\documentclass[prl,twocolumn,amsmath,amssymb,floatfix,footinbib,bibnotes,longbibliography]{revtex4-2}
\pdfoutput=1

\usepackage[colorlinks=true,citecolor=blue,linkcolor=magenta]{hyperref}
\usepackage{multirow}

\usepackage{amsmath}
\usepackage{bbm}
\usepackage{graphicx}
\usepackage{dsfont}
\usepackage{changepage}
\usepackage[normalem]{ulem}
\usepackage{fancyhdr}
\usepackage{amsthm, amssymb}
\usepackage{array}
\newcolumntype{P}[1]{>{\centering\arraybackslash}p{#1}}
\usepackage[usenames,dvipsnames]{color}
\usepackage[section]{placeins}
\usepackage[T1]{fontenc}
\usepackage[active]{srcltx}
\usepackage{color}
\usepackage{amssymb}
\usepackage{esint}
\usepackage[version=4]{mhchem}
\usepackage{mathtools,mathrsfs}

\newcommand{\beginsupplement}{%
        \setcounter{table}{0}
        \renewcommand{\thetable}{S\arabic{table}}%
        \setcounter{figure}{0}
        \renewcommand{\thefigure}{S\arabic{figure}}%
}

\newcommand {\apgt} {\ {\raise-.5ex\hbox{$\buildrel>\over\sim$}}\ }
\newcommand {\aplt} {\ {\raise-.5ex\hbox{$\buildrel<\over\sim$}}\ }

\newcommand{\red}{\textcolor{red}}

\def\titlename{Symmetry, Superposition and Fragmentation in Classical Spin Liquids: A General Framework and Applications to Square Kagome Magnets}

\def \authornames{K.B. Yogendra, Suman Karmakar, and Tanmoy Das}
\def \affiliations{Department of Physics, Indian Institute of Science, Bangalore 560012, India \\}

\begin{document}

\title{\titlename}
\author{\authornames}
\affiliation{\affiliations}

\begin{abstract}
Classical magnets exhibit exotic ground state properties such as spin liquids and fractionalization, promising a manifestation of superposition and projective symmetry construction in classical theory. While system-specific spin-ice or soft-spin models exist, a formal theory for general classical magnets remains elusive. Here, we introduce a generic symmetry group construction built from a vector field in a plaquette of classical spins, demonstrating how classical spins superpose in irreducible representations (irreps) of the symmetry group. The corresponding probability amplitudes serve as order parameters and local spins as fragmented excitations. The formalism offers a many-body vector field representation of diverse ground states, including spin liquids and fragmented phases described as degenerate ensembles of irreps. We apply the theory to a frustrated square Kagome lattice, where spin-ice or soft spin rules are inapt, to describe spin liquids and fragmented phases, all validated through irreps ensembles and unbiased Monte Carlo simulation. Our generic theory sheds light on previously unknown aspects of spin-liquid phases and fragmentation and broadens their applications to other branches of field theory. 
\end{abstract}

\maketitle


Classical spin models can potentially capture exotic phenomena like spin liquid \cite{balents2010spin, Henley2010Review, MoessnerCSLpyro98, Rehn2016CSL_Honeycomb, BentonPyroclore16, Benton2021}, spin ice \cite{Castelnovo2012, Nisoli2013IceReview, Ross2011QIce}, and fragmentation \cite{Henley2010Review,Brooks2014Peter,Powell2015,petit2016fragm,Canals2016fragm}, order by disorder \cite{Villain,Henley,Chandra,Chalker,Shenderbook,Biskup_2004}, prethermal discrete time crystals\cite{Knolle2021PrethermalDTC}, and exciting progress lies in designing novel and generic frameworks \cite{Benton2021,Placke2023ising,DahlbomBatista23,Rajah2023, Davier2023,Yan2023Clas,Yan2023Clas_Detailed,Fang2023Clas,Zohar_2004,Zohar_2004Comm}. While quantum theory allows the ground state of a spin liquid to be a superposition state, this concept does not have a classical analog. 
Classically, two main approaches so far describe the spin liquid phase.  The spin-ice applies to specific spin Hamiltonians that can be expressed in terms of $|{\bf S}_{c}|^2$, where the total spin in a cluster $c$: ${\bf S}_c=\sum_{i\in c}\eta_i{\bf S}_i$ with $\eta_i$ being suitably chosen rational numbers\cite{Benton2021}. This way the  ${\bf S}_c=0$  configuration describes a degenerate ground state.\cite{MoessnerCSLpyro98,Henley2010Review,Benton2021,Powell2015,Rajah2023,Davier2023}  However, this rule doesn't hold for models with Dzyaloshinskii-Moriya (DM) interactions. Recently, a Luttinger-Tisza approximation, also known as the spherical or soft-spin approximation, has been employed to analyze the degenerate energy state in momentum space in terms of extended states of classical spin.\cite{Rajah2023, Davier2023, Yan2023Clas, Yan2023Clas_Detailed, Fang2023Clas}
A flat band in this model indicates the degeneracy characteristic of spin liquids. The drawback of this model is that it relaxes the local $|{\bf S}_i|=1$ constraint, imposing it at the global spin value. Both approaches are suited for specific Hamiltonians and have so far been applied only to spin-ice models.

Magnetic fragmentation is another exotic phenomenon in the classical spin systems that draws recent attention.\cite{Bramwell,Brooks2014Peter,Canals2016fragm,petit2016fragm,Powell2015,Savary2012} In this phase, a local classical field (such as spin or magnetization) fragments into components with one (or more) components exhibiting order while others remain disordered or liquid-like. This phenomenon has so far been studied using Landau's coarse-grained magnetization fields, with or without local spin constraints. Despite progress in understanding specific models with ground state degeneracy or fragmentation, a comprehensive analytical framework, which would ideally encompass all lattice symmetries, frustration, DM interactions, local spin constraints, and hence do not necessarily follow the spin-ice rule, remains elusive.

Research on frustrated lattices, like pyrochlore,\cite{Siddharthan1999Pyrochlore, Bramwell2001Pyrochlore, Isakov2004,Yan2017,Benton2021} triangular,\cite{Mol2012ArtificialTIce, Li_2020Triangular, Liu2016Triangular} Kagome,\cite{Yasir2015Kagome, Mizoguchi2017kagome, Essafi2017} and others\cite{Henley,Yildirim,Shenderbook,Rehn2016CSL_Honeycomb} has been a major focus in exploring spin liquids and related phenomena. Recently, the square Kagome lattice has sparked excitement due to experimental hints of spin liquid phases\cite{fujihala2020, Yakubovich2021,markina2023} and supporting theoretical investigations \cite{Yasir2021, Richter2022,Niggemann2023, Richter2023, Gembe2023}. However, these materials likely possess a strong DM interaction\cite{fujihala2020, Yakubovich2021,markina2023} which the spin-ice and soft-spin models do not incorporate. Additionally, the square Kagome lattice boasts multiple sublattices, offering a richer platform with potentially larger degenerate manifolds and more fragmentation possibilities. 

Here, we introduce a generic framework for studying ground state degeneracy and fragmentation in classical spin systems using a group theory approach. We apply this theory to a two-dimensional square Kagome lattice. Our approach transcends a prior approach\cite{Yan2017, Essafi2017}, used primarily for ordered phases, to encompass spin liquids and fragmented phases. We define a vector space representing the spins within a lattice plaquette, designed to be invariant under the lattice's point-group symmetry.  The plaquette spin vector can be expressed as a superposition of the irreducible representations (irreps) of the symmetry group. The expansion parameters of this superposition vector act as Landau-like order parameters. However, unlike traditional order parameters, they transform under `discrete' spatial rotations and acquire continuous symmetry through degeneracy and irreps multiplets. Interestingly, these order parameters serve as spin's `probability amplitudes'  and `occupation densities' to irreps state and energy levels, respectively. This approach, with its resemblance to quantum concepts, paves the way for a novel construction of classical spin liquids and fragmentation states. 

We apply the theory to a model consisting of  XXZ and DM interactions in a 2D square Kagome lattice. We also employ unbiased classical Monte Carlo (MC) simulation to validate our group theory approach and reproduce the phase diagram. We find that DM interaction promotes a uniform or staggered ordering of specific irrep, containing vortex or anti-vortex. Near the critical boundaries between these ordered phases, we observe the emergence of classical spin liquid (CSL) states. Within the CSL phase, local spins remain fully disordered if the ground state consists of a randomly distributed irrep ensemble. Alternatively,  the ground state can scramble ordered and disordered irreps to fragment the local spin vector into coexisting extended/collective and point-like excitations. Additionally, the spin-spin correlation function is analyzed in each phase to distinguish between magnetic Bragg peaks for the collective excitations in the ordered phase and the `pinch-point' excitations in the liquid phases. 

\begin{figure}[tb!]
	\centering
	\includegraphics[width=0.99\linewidth]{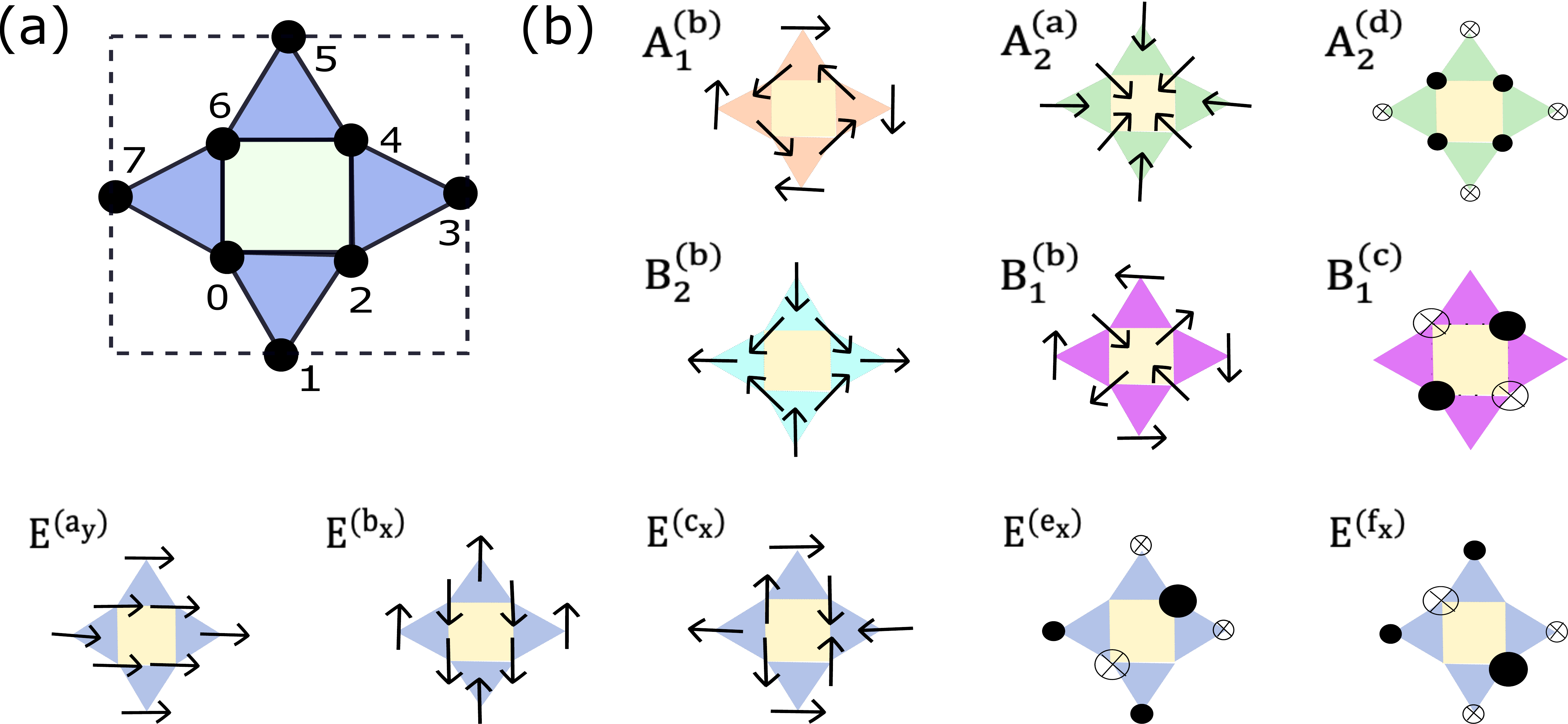}
	\caption{(a) A plaquette of a 2D square-Kagome lattice, belonging to the $\mathsf{D}_4$ group, is shown with sublattices enumerated as $i=0-7$. (b) Among five irreps with different multiplets, we show a few representative irreps here, while others are shown in \textit{SM}\cite{SM}. Each irrep consists of either $S_i^{\perp}$ (horizontal arrow) or $S_i^z$ (open and filled dots for up and down spins) components, with the sizes of the arrows or dots dictate their magnitudes. }
    \label{fig:symLattice}
\end{figure}

\textit{Mathematical foundation:} 
We define a local vector field in a plaquette network  $p$ to be invariant under the lattice's point-group symmetry $\mathsf{G}$:
\begin{equation}
\mathcal{S}_p = \bigoplus_{i\in p} S_i.
\label{Eq:DefSp}
\end{equation}
$S_i=(S^x_i~S^y_i~S^z_i)^{T}\in \mathsf{O}_i(3)$ at the $i^{\rm th}$ site, and  $\mathcal{S}_p\in \mathsf{O}_p(3n)$ where $n$ is the number of sublattices in $p$. ($\mathsf{O}_i(n)$, $\mathsf{O}_p(n)$ distinguish the orthogonal symmetry of the vector at the $i$-site and $p$ - plaquette, respectively). Each plaquette, like a conventional unit cell, includes redundant sites than the primitive unit cell. This is adjusted by introducing a normalization factor in the dual vector definition to fix the length of $\mathcal{S}_p$.\cite{dualspace}

The transformation from the spin space to the irreps space of group $\mathsf{G}$ involves an orthogonal matrix, whose column vectors $\mathcal{V}_{\alpha}$ form the orthonormal basis of the irreps representation. Expressing $\mathcal{S}_p$ in this irreps space yields
\begin{equation}
\mathcal{S}_p =\sum_{\alpha=1}^{3n}m_{\alpha}\mathcal{V}_{\alpha}.
\label{Ex:ExpandSp}
\end{equation}
Here $m_{\alpha}\in \mathbb{R}$ is the coefficient of the expansions. We keep the plaquette index in $m$ and $\mathcal{V}$ implicit for simplicity in notation. Interestingly, $m_{\alpha}$ conforms to Landau's order parameter as the coarse-grain average of local fields, except, here, it is invariant under a discrete symmetry group in a plaquette and is interpreted as the probability amplitude of vector field:  $m_{\alpha}=\mathcal{V}_{\alpha}^{\mathcal{T}}\mathcal{S}_p$.\cite{dualspace} The local spins are defined by a rectangular projection matrix $\mathcal{P}_{i\in p}$ as ${\bf S}_{i\in p}=\mathcal{P}_{i\in p}\mathcal{S}_{p}=\sum_{\alpha}{m}_{\alpha}\mathcal{P}_{i\in p}\mathcal{V}_{\alpha}$. 

Reformulating the order parameters in terms of the irreps conveniently decouples them in a symmetry invariant Hamiltonian, albeit the irreps' multiples can mix among themselves.  To account for the multiplets' submanifold and emergent symmetry, it is convenient to introduce an $\mathsf{O}_p(d_{\alpha})$ `spinor'-like field $\boldsymbol{m}_{\alpha}:=(m_{\alpha}^{(1)}~ ...~m_{\alpha}^{(d_{\alpha})})^{T}$ for the $\alpha$ irrep with $d_{\alpha}$ multiplet. Then, the eigenmodes are obtained by orthogonal rotation $\tilde{{\boldsymbol{m}}}_{\alpha}=e^{i\boldsymbol{\mathcal{L}}_{\alpha}\cdot{\boldsymbol{\phi}}_{\alpha}}\boldsymbol{{m}}_{\alpha}$, where $\boldsymbol{\mathcal{L}}_{\alpha}$ are the corresponding generators for the angle $\boldsymbol{\phi}_{\alpha}$. $\boldsymbol{\phi}_{\alpha}$ lives on the Hamiltonian's parameter space and assumes fixed values for the energy eigenmodes. The orthonormal basis states ensure the constraint $|\mathcal{S}_p|^2=\mathcal{S}_p^{\mathcal{T}}\mathcal{S}_p=\sum_{\alpha}d_{\alpha}|m_{\alpha}|^2=nS^2$, $\forall p$, where $|S_i|=S$, $\forall i$ is an additional hardcore constraint on the classical spins\cite{dualspace}. Not all irreps necessarily adhere to the local constraint, requiring them to collaborate with others for existence. Such irreps ensembles may lead to non-analyticity and fragmentation into an order-disorder mixed phase. Additionally, the collapse of the eigenmodes $\tilde{{\boldsymbol{m}}}_{\alpha}$ into its constituent irrep $\boldsymbol{{m}}_{\alpha}$ causes distinct fragmented excitation.

We have a $3nN$-dimensional vector space $\mathcal{S}=\bigoplus_p \mathcal{S}_p$ for a generic $N$-unit cell lattice, commencing a $3nN\times3nN$-matrix valued quadratic-in-spin Hamiltonian (see SM\cite{SM} for further details). However, thanks to nearest-neighbor interaction and discrete-translation-invariance of the lattice, the Hamiltonian can be brought to a block-diagonal form in terms of the plaquette Hamiltonian $H_p$:
\begin{equation}\label{eq:Hp}
    H_p= \frac{1}{2}\mathcal{S}_p^{\mathcal{T}}\mathcal{H}_p\mathcal{S}_p.
\end{equation}
Here $\mathcal{H}_p$ is an orthogonal matrix-valued Hamiltonian, analogous to the second quantized Hamiltonian, whose components consist of all possible interactions between ${\bf S}_i$ and ${\bf S}_j$ for $\langle ij\rangle\in p$. However, lattice symmetries restrict the allowed finite components in $\mathcal{H}_p$, which we now consider for a square kagome lattice. 

{\it Realizations in a square-Kagome lattice:} The square-Kagome lattice belongs to the Dihedral ($\mathsf{D}_4$) group with $n=8$ sublattice spins, giving a $24$-dimensional vector representation. Denoting the group element $\mathsf{g}\in\mathsf{D}_4$ in the $\mathcal{S}_p-$representation by the matrix-valued operators $\mathcal{D}(\mathsf{g})$, we impose the symmetry criterion that under a local symmetry transformation $\mathcal{S}_p\rightarrow \mathcal{D}(\mathsf{g})\mathcal{S}_p$, the local Hamiltonian $H_p$ is invariant if $[\mathcal{D}(\mathsf{g}),\mathcal{H}_p]=0$, $\forall p,\mathsf{g}$. Since local $\mathsf{O}_i(3)$ and sublattice symmetries are abandoned, the plaquette symmetry allows us to have bond- and spin-dependent interactions $J^{\mu\nu}_{ij}$ with six exchange and three DM interactions (see \textit{SM} for the details), leading to a bond-dependent XYZ-Heisenberg model with XY-DM interaction. However, imposing bond-independent interactions, we consider an XXZ model with DM interaction as more appropriate for realistic materials \cite{fujihala2020, Yakubovich2021,markina2023}, $H =     \sum_{\langle ij\rangle,\mu\nu}J^{\mu\nu} S_i^{\mu} S_j^{\nu}$. This can, for future convenience, be expressed as:  
\begin{eqnarray}{\label{eq:Hamxxz}}
    H    = J\sum_{\langle ij\rangle,\tau=\pm}\left(D^{\tau}e^{ {\rm i}\tau(\Theta_i+\Theta_j)}S^{\perp}_iS^{\perp}_j +\Delta S_i^z S_j^z \right).
\end{eqnarray}
Here $J^{\mu\nu}=J\delta_{\mu\nu}+JD\epsilon_{\mu\nu}$ for $\mu=x,y$, and $J^{zz}=J\Delta $, $\delta_{\mu\nu}$ is the Kronecker delta and $\epsilon_{\mu\nu}$ is the Levi-Civita tensor. $J$ is the exchange term, $\Delta$ is the $z$-axis anisotropy ratio, and $JD$ is the XY DM interaction strength. By diagonalizing the tensor $J^{\mu\nu}$, we define two `circularly polarized' fields: $S_i^{\tau}=|S^{\perp}_i|e^{ {\rm i}\tau\Theta_i}\in \mathsf{O}_i(2)\cong \mathsf{U}_i(1)$, where $S^{\perp}_i=\sqrt{S^2-(S^z_i)^2}$ is the coplanar spin magnitude and $\Theta_i$ is the azimuthal angle in the spin space, which interact via a complex (dimensionless) interaction $D^{\tau}=1+{\rm i}\tau D$.

\begin{figure*}[t]
    \centering
    \includegraphics[height=0.280\linewidth,width=1.0\linewidth]{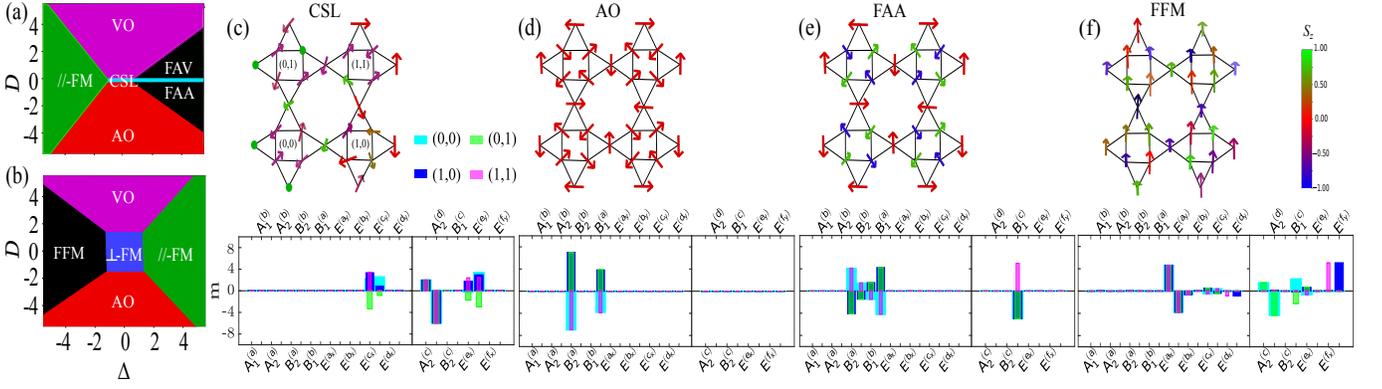}
    \caption{Computed phase diagrams within the MC simulation (also group theory analysis) are shown for (a) for AFM ($J=+1$) and (b) for the FM ($J=-1$) couplings. We highlight spin textures in a randomly chosen four-plaquette setting for representative phases (upper panel) and respective ensembles of irreps in four plaquettes (lower panel). (c) CSL at $(J,\Delta, D)=(1,1,0)$ showing disordered values of $m_{\nu}$ from both in-plane and out-of-plane ensambles. (d) AO at $(1,0,-3)$ with degenerate irreps $\mathsf{B}_{1,2}^{\rm (a)}$  are staggered. (e) FAA phase at $(1,4,-1)$ where $\mathsf{B}_{1,2}^{\rm (a)}$ being ordered but $\mathsf{B}_{1}^{\rm (c)}$ is disordered. (f) FFM phase at $(-1,-2.5,0)$ where 2D irrep $\mathsf{E}^{\rm(a)}$ is ferromagnetically ordered in-plane, but out-of-plane irreps are disordered. Note that all disordered values take random numbers between different plaquettes, while we display only four representative plaquettes here.}
    \label{fig:xxzPhase}
\end{figure*}

{\it Irreps in square-Kagome lattice:} There are five conjugacy classes in this non-Abelian group, giving five irreps: $m_{\alpha}\equiv$ $\mathsf{A}^{(d_{\alpha})}_{1,2}$, $\mathsf{B}^{(d_{\alpha})}_{1,2}$, and one two-dimensional irrep $\mathsf{E}^{(d_{\alpha})}$, where the superscript denotes their multiplicity $(d_{\alpha})=(2,4,3,3,6)$, respectively.  Representative irreps configurations are shown in Fig.~\ref{fig:symLattice}(b).  We organize these irreps into a coplanar set $\mathsf{m}_{\perp}:=\{\mathsf{A}_{1,2}^{\rm(a,b)}, \mathsf{B}_{1,2}^{\rm(a,b)}, \mathsf{E}^{\rm(a_{x,y},b_{x,y},c_{x,y},d_{x,y})}\}$, and an  out-of-plane/colinear set $\mathsf{m}_{||}:=\{\mathsf{A}_2^{\rm(c,d)}, \mathsf{B}_{1,2}^{\rm (c)}, \mathsf{E}^{\rm (e_{x,y},f_{x,y})}\}$. 

In the coplanar irreps basis $\mathsf{A}^{\rm{(a,b)}}_{1,2}$, $\mathsf{B}_{1,2}^{\rm{(a,b)}}$, even/odd under $\mathsf{C}_4$, the local spins $S_i^{\tau}$ are arranged in a topological texture following $\Theta_{i\in p}= Q_p \theta_i+\gamma_p$, where $\Theta_i$ and $\theta_i$ are the azimuthal angles in the spin and position manifolds, respectively. $\gamma_p\in [0,\pi)$ is the helicity angle, and $Q_p\in \pi_1(\mathbb{S}^1)\cong\mathbb{Z}$ is the topological charge. As shown in Fig.~\ref{fig:symLattice}(b), this leads to (anti-/) vorties for $\mathsf{A}^{\rm{(a,b)}}_{1,2}$, $\mathsf{B}_{1,2}^{\rm{(a,b)}}$ irreps.   In fact, each (anti-/) vortex consists of two concentric (anti-/) vortices in the outer and inner squares, which are not related by symmetry but interact with each other by the interaction term $D^{\tau}$. $\mathsf{A}_1^{\rm{(a,b)}}$ consist of concentric vortices with the same/opposite helicities ($\gamma_p=\pm \pi/2$), while $\mathsf{A}_2^{\rm{(a,b)}}$, odd under reflection, have $\gamma_p=\pm \pi$. $\mathsf{B}_{1,2}^{\rm{(a,b)}}$ irreps (odd under $\mathcal{C}_{4}$) are similar, except they contain anti-vortices. The out-of-plane  $\mathsf{A}_2^{\rm{(c,d)}}$ are colinear FM/AFM irreps, while $\mathsf{B}_{1,2}^{\rm{(c)}}$ are colinear AFM irreps. Finally, among the six-fold multiplets of 2D $\mathsf{E}$ irreps, $\mathsf{E}^{\rm (a-d)}$ are coplaner FM/ nematic/AFM order parameters, while $\mathsf{E}^{\rm (e,f)}$ are colinear irreps. Notably, the colinear irreps $\mathsf{B}_{1,2}^{\rm (c)}$ and $\mathsf{E}^{\rm (e,f)}$ violate the local constraints, and hence their low-energy configurations vitiate any long-range order.  

{\it Eigen energies:} The final task is to diagonalize the multiples of the irreps. In our case, the irreps' multiplets split as either $\mathsf{O}_p(d_{\alpha})=\mathsf{O}_p(2)\oplus\mathsf{O}_p(2)\oplus ...$, or $\mathsf{O}_p(d_{\alpha})=\mathsf{O}_p(2)\oplus\mathsf{Z}_2\oplus ...$, in which all $\mathsf{O}_p(2)$ operators have the same generator $\mathcal{L}_{\alpha}=i\sigma_y$. $\phi_{\alpha}$ depends only on $\rm {arg}(D^{\tau})$ in the eigenstates of $\mathcal{H}_p$. The resultant diagonal Hamiltonian per plaquette is
\begin{eqnarray}
H_p=\sum_{\nu=1}^{3n}E_{\nu}|\tilde{{\bf m}}_{\nu}|^2.
\label{eq:eigenE}
\end{eqnarray}
Here $|\tilde{{\bf m}}_{\nu}|^2$ serves as `occupation density' to the ${\nu}^{\rm th}$ energy level $E_{\nu}$. Henceforth, we omit the tilde symbol for simplicity, and all irreps are considered eigenmodes unless mentioned otherwise.  The functional form of $E_{\nu}$ in terms of $J$, $D$, and $\Delta$ is given in the SM\cite{SM}. Constrained by symmetry, $E_{\nu\in \mathsf{m}_{\perp}}$ depends solely on $D^{\tau}$, while $E_{\nu\in \mathsf{m}_{||}}$ is proportional to $\Delta$ \footnote{Specifically, the (anti-/) vortex irreps $\mathsf{B}_{1,2}^{\rm{(a,b)}}$, $\mathsf{A}_{1,2}^{\rm{(a,b)}}$ are promoted by $\mp D$, while $\mathsf{E}^{\rm(a,b)}$ do not depend on $D$.}. One or more irrep (s) can form an ordered phase with a global energy minimum at $NE_{\nu}$ if they satisfy the constraint and frustration; otherwise, they blend with other irreps to form a degenerate ensemble, giving disorder, liquid, and mixed phases. A zero-temperature phase transition occurs at the $E_{\nu}=0$ line.

\begin{table*}
\centering
\begin{tabular}{| c| c| c| c|c|}
 \hline
 Phase & Acronym & Irreps $\{\nu_p\}$ & Parameters & Color code \\ 
 \hline
 Classical Spin Liquid & CSL & $\mathsf{m}_{\perp}\cup \mathsf{m}_{||}$ & $J=1$, $\Delta>0$, $D=0$& Cyan \\  
 \hline
 Vortex Order & VO  & ${\mathsf{\bar{A}}_{1,2}^{\rm (a)}}$& $J=1$, $\Delta<2D$, $D>0$& Magenta \\  
 \hline
 Anti-vortex Order & AO & $\mathsf{\bar{B}}_{1,2}^{\rm (a)}$& $J=1$, $\Delta<2D$, $D<0$& Red \\  
 \hline
 Fragmented AFM-vortex  & FAV & $\mathsf{\bar{A}}_{1,2}^{\rm (a)} \cup \mathsf{B}_1^{\rm (c)}$& $J=1$, $\Delta>2D$, $D>0$& Black \\
 \hline
 Fragmented AFM-Anti-vortex & FAA & $\mathsf{\bar{B}}_{1,2}^{\rm (a)} \cup \mathsf{B}_1^{\rm (c)}$& $J=1$, $\Delta>2D$, $D<0$& Black \\
 \hline
 Fragmented Ferromagnet & FFM  & $\mathsf{\bar{E}}^{\rm{(a)}}\cup \mathsf{m}_{||}$& $J=-1$, $\Delta>2|D|$,  $\pm D$& Black \\
 \hline
 Colinear Ferromagnet Order& $||$-FM& $\mathsf{\bar{A}}_2^{\rm (c)}$& $\Delta J<0$,  $|\Delta|>2|D|$& Green \\
 \hline
 Coplanar Ferromagnet Order& $\perp$-FM & ${\mathsf{\bar{E}}^{\rm (a)}}$& $J=-1$,  $|\Delta|<2|D|$ & Blue \\
 \hline
   \end{tabular}
   \label{tab:Phase}
   \caption{We tabulate all the phases and the contributing irreps obtained consistently with the MC simulation and the group theory analysis. The irrep with a bar in the third column reflects it to be ordered; otherwise, it's a disorder irrep.}
\end{table*}

{\it Phase diagrams and correlation functions:} We solve the Hamiltonian in Eq.~\eqref{eq:Hamxxz} both numerically using classical MC simulations and the group theory analysis. The details of the MC simulation are given in the SM\cite{SM}. The corresponding phase diagram is summarized in Table~1 and shown in Fig.~\ref{fig:xxzPhase}. Note that the same phase diagram is also reproduced by the lowest energy eigenvalue $E_{\nu}$, and the values of $m_{\nu}$ are obtained from the MC result as shown in the lower panel in  Fig.~\eqref{fig:xxzPhase} agrees with the group theory result.

Remarkably, we find that all the phases can be understood in terms of an analytical definition of the many-body ground state vector field as:
\begin{eqnarray}
\mathcal{S}_{\rm GS} &=& \bigoplus_p\sum_{\{\nu_p\}} m_{\nu_p}\mathcal{V}_{\nu_p}.
\end{eqnarray}
The ordered phase harbors a summated state of a fixed irrep $\bar{\nu}\in \{\nu_p\}$ (with $m_{\bar{\nu}}=\bar{m}$, $m_{\nu\ne\bar{\nu}}=0$, $\forall p$); while the staggered phase features two alternating but fixed irreps $\bar{\nu}_p$ and $\bar{\nu}_q$ in neighboring plaquettes. The CSL state, on the other hand, combines an ensemble of irreps $\{\nu_p\}$ within each plaquette $p$. Within this ensemble, the probability amplitude $m_{\nu_p}$ may vary randomly, subject to local constraints, corresponding to the same plaquette energy. The random distribution of $m_{\nu_p}$ differs between plaquettes, resulting in an extensively degenerate ground state. 

In addition, we also compare our results with a soft-spin approximation in the Fourier space \cite{Isakov2004, Conlon2009, Conlon2010, Rehn2016CSL_Honeycomb, Rehn2017CHeisenberg, Benton2021}, and the resulting dispersion relation is shown in \textit{SM}\cite{SM}. Given that we have experimental access to the correlation function of local spins ${\bf S}_{i\in p}$, we report its correlation function.  
We project the structure factor $
    \chi(\textbf{k}) = 1/\mathcal{N} \sum_{i,j} \langle \textbf{S}_i \cdot \textbf{S}_j \rangle \exp{({{\rm i}\textbf{k} \cdot (\textbf{r}_i - \textbf{r}_j)})}$ to the irreps space as  
    \begin{eqnarray}
    \langle \textbf{S}_i \cdot \textbf{S}_j \rangle &=& \sum_{\nu_p\nu_q}{m}_{\nu_p}{m}_{\nu_{q}} \langle \mathcal{{V}}_{\nu_p}^{\mathcal{T}} \mathcal{P}^T_{i}\mathcal{P}_{j}\mathcal{{V}}_{\nu_{q}}\rangle, 
\end{eqnarray}
with \textbf{r}$_i$ is the $i^{\rm th}$ spin's position in $p$ and $j\in q$ plaquette.

\begin{figure}[t]
    \centering
    \includegraphics[width=1.0\linewidth]{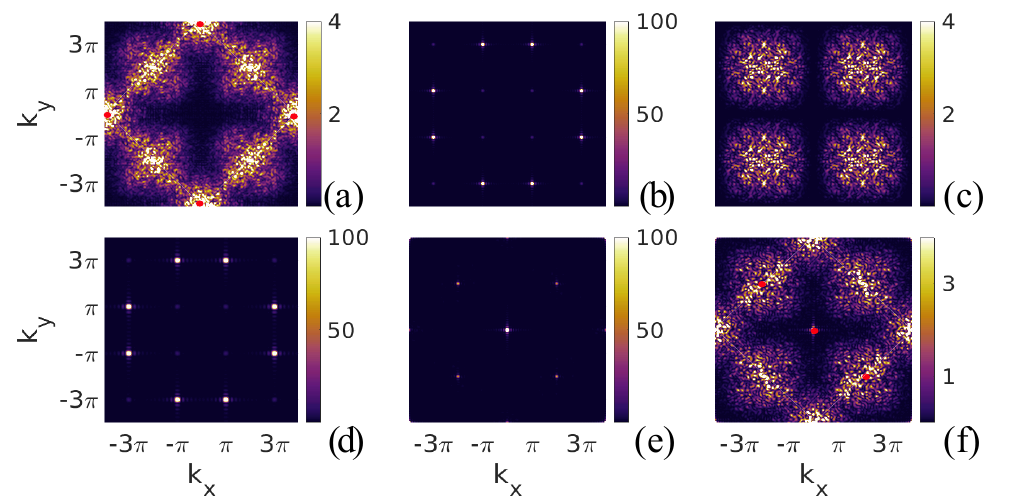}
    \caption{Simulated $\chi({\bf k})$ is plotted in the momentum space for the four phases discussed in Fig.~2. (a) CSL at $(J,\Delta, D)=(1,1,0)$, where red dots are plotted separately to signify additional strong magnetic Bragg-like peaks that overwhelm the spectral density of the disordered pattern. (b-c) FFA at $(1,4,-1)$ where the plots for the ordered $S_i^{\perp}$ and disordered $S_i^z$ components are separated in (b) and (c), respectively. (d) AO at $(1,0,-3)$ showing Bragg peaks similar to $S_i^{\perp}$ components in (b). (e-f) FFM phase at $(-1,-2.5,0)$ with FM ordered $S_i^{\perp}$ and disorder $S_i^z$ are separated in (e) and (f). Panels (a) and (f) host pinch-points around $(\pi,3\pi)$ and it;s equivalent points.}
    \label{fig:MixedStr}
  \end{figure}
  
The phase diagram in Fig.~\ref{fig:xxzPhase} reveals a predominance of (uniform or staggered) order phases in both $J<0$ (frustration inactive) and $J>0$ (frustration active) regions. A CSL phase emerges only at the critical line of $D\rightarrow 0$, which turns into distinct mixed/fragmented phases for $2|D|/\Delta<1$. For $D\rightarrow 0$, $J>0$, three distinct CSL phases emerge with increasing $\Delta$ in Fig.~\ref{fig:xxzPhase}(a) (cyan color). As $\Delta\rightarrow 0$, we have an XX model in Eq.~\eqref{eq:Hamxxz}, and the contributing irreps arise from the degenerate manifold of the coplanar irrep ensemble $\{m_{\nu_p}\}\subseteq \mathsf{m}_{\perp}$. This gives a CSL phase of $S_i^{\tau}\in \mathsf{O}_i(2)$ spins.  The structure factor $\chi({\bf k})$ displays a characteristic disorder pattern without any magnetic Bragg peak but with a prominent pinch-point around ${\bf k}=(\pm\pi,\pm3\pi)$. The pinch-point characterizes a Coloumb phase with algebraic spin-spin correlation feature\cite{Henley2010Review}. At $\Delta=1$, the Hamiltonian is subject to a full $\mathsf{O}_i(3)$ symmetry constraint per site, resulting in symmetry-allowed access to the entire ensemble $\{m_{\nu_p}\}\subseteq \mathsf{m}_{\perp}\cup \mathsf{m}_{||}$. For example, $\{m_{\nu}\}\in\{m_{\mathsf{A}_{1,2}^{\rm(a,b,c,d)}}, m_{\mathsf{B}_{1,2}^{\rm(a,b)}}\}$ are degenerate at $E_{\nu}=-2J$ and $\{m_{\nu'}\}\in\{m_{\mathsf{B}_{1,2}^{\rm(c)}}, m_{\mathsf{E}^{\rm(c,d)}}\}$ at $E_{\nu'}=-4J$, making a larger CSL ensemble degenerate at energy $E_p=m_{\nu}^2E_{\nu}+m_{\nu'}^2E_{\nu'}=-4J$  for $m_{\nu}=\sqrt{2}m_{\nu'}$. Consequently, $\chi({\bf k})$ displays pinch-point correlations among both $S_i^{\tau}$ and $S_i^z$. Finally, as $\Delta\rightarrow \infty$, the Hamiltonian (last term in Eq. \eqref{eq:Hamxxz}) retains a residual local $\mathsf{Z}_2$ symmetry constraint, and the disorder ground state solely stems from the $\{m_{\nu_p}\}\subseteq \mathsf{m}_{||}$ ensemble. $\chi({\bf k})$ is contributed solely by $S_i^z$ with pinch-points at $\bf k=(\pm\pi,\pm3\pi)$. Based on their distinct local constraints, it is convenient to refer to these phases as $\mathsf{O}(2)$, $\mathsf{O}(3)$, and $\mathsf{Z}_2$ CSLs, respectively, without implying a Landau-type phase boundary between them. 

Any finite $D$ steers the CSL phase into either order or fragmented (mixed) phases. Note that vortex irreps $\mathsf{A}_{1}^{\rm {(a,b)}}$ and $\mathsf{A}_{2}^{\rm {(a,b)}}$  are degenerate at $E_{\nu}=2D\pm 2\sqrt{D^2+(1+D)^2}$, while the anti-vortex irreps $\mathsf{B}_{1}^{\rm {(a,b)}}$ and $\mathsf{B}_{2}^{\rm {(a,b)}}$ are degenerate at $E_{\nu}=-2D\pm 2\sqrt{D^2+(1-D)^2}$. This makes all the phases in Fig.~\ref{fig:xxzPhase}(a),(b) symmetric for $D\leftrightarrow -D$ with vortices $\leftrightarrow$ ant-vortices. Hence, we mainly focus on the $-D$ region with anti-vortices for the discussions.

For weak out-of-plane anisotropy $\Delta< 2|D|$,  we have ordered phases of (anti-/) vortices for $\mp D$, which we call (Anti-/)Vortex Order (AO/VO) phases (red/magenta regions in Fig.~\ref{fig:xxzPhase}). In AO phase, the degenerate irreps $\mathsf{B}_{1,2}^{\rm {(a)}}$ are mixed in an $\mathsf{O}(2)$ order parameter and are staggered between the neighboring plaquettes with a $\gamma_p=\pi$ phase shift. The extracted values of the order parameter $m$ from the MC data confirm the only finite and uniform weight of the $\bar{m}_{\mathsf{B}_{1,2}^{\rm (a)}}$ irreps in the AO phase, as shown in Fig.~\ref{fig:xxzPhase}(d) (lower panel). Interestingly, the CSL lies at the phase transition line between the VO and AO phases. The ordering is also evident in $\chi({\bf k})$ with a magnetic Bragg peak at ${\bf k}=(\pi,\pi)$. 

However, for strong $\Delta> 2|D|$ (with AFM anisotropy $J\Delta>0$ ), the coplanar ordered irreps become scrambled with disordered out-of-plane irreps: $\{m_{\nu_p}\}_{\rm mix}\subseteq \bar{m}_{\mathsf{A}/\mathsf{B}}\cup \mathsf{m}_{||}$, in black region Fig.~\ref{fig:xxzPhase}(a). In particular, the outer (anti-/) vortex maintains co-planarity, while the inner (anti-/) vortex mixes with the $m_{\mathsf{B}_1^{{\rm(c)}}}\in\mathsf{m}_{||}$ irrep. The combination produces a novel \textit{AFM-vortex/AFM-anti-vortex} texture within the inner square where neighboring spins possess opposite easy axes \footnote{This AFM-vortex topology is homotopically distinct from the known AFM skyrmion\cite{AFMskyrmion}, and has not been predicted previously}. Consequently, ${\bf S}_i$ spin fragments into its $S_i^z$ components become non-interacting and fail to order or exhibit any significant correlation, while the $S_i^{\tau}$ fields exhibit long-range order with magnetic Bragg peaks in the structure factor, see Figs.~\ref{fig:MixedStr}(b,c). We denote these phases as fragmented AFM-vortex (FAV) and fragmented AFM-Anti-vortex (FAA) for $\pm D$ regions and confirm the same the extracted values of $m_{\nu}$ from the MC result.

For strong $\Delta> 2|D|$ with FM anisotropy $\Delta<0$ and $J>0$ naturally select colinear FM order of the $\mathsf{A}_2^{\rm (c)}$ irrep (green region Fig. ~\ref{fig:xxzPhase}(a)). We denote this phase as $||-$FM. The same phase reemerges for $\Delta>0$ and $J<0$ in  Fig. ~\ref{fig:xxzPhase}(b).

The interplay between the FM interaction, $J=-1$, and strong AFM anisotropy $\Delta>2|D|$ generates a distinct fragmented phase, see Fig.~~\ref{fig:symLattice}(b) (black region).  The extracted values of $m$ from the MC data show that the in-plane FM 2D irrep $\bar{m}_{\mathsf{E}^{\rm{(a)}}}$ is ordered while the out-of-plane AFM irreps $\in m_{||}$ are disordered, see Fig.~~\ref{fig:symLattice}(f). These out-of-plane irreps violate the local constraint, leading to an intriguing fragmented structure in $\chi({\bf k})$, resulting in an in-plane FM order in $S_i^{\tau}$, but a pinch-point disorder in $S^z_i$, see Fig.~\ref{fig:MixedStr}(e-f). We dub this a Fragmented FM (FFM) phase.

Any finite $D$ disfavors this mixed phase, causing a phase transition at $D>2\Delta$ to in-plane VO or AO orders for $\pm D$, as observed in the $J=1$ phase diagram. The remaining two phases are readily identifiable: a uniform coplanar FM (namely, $\perp$-FM) order with $\bar{m}_{\mathsf{E}^{\rm (a)}}$ irrep at $\Delta\rightarrow 0$ (blue region in Fig.~\ref{fig:xxzPhase}(b)), and an out-of-plane $||$-FM order with $\bar{m}_{\mathsf{A}_2^{\rm (c)}}$  for  $J\Delta\rightarrow \infty$ (green region in Fig.~\ref{fig:xxzPhase}(b)).

\textit{Conclusions and outlook.} Discussions on their excitations and phase transition are merited. The VO/AO order phases (red and magenta) exhibit novel collective excitations. Gapless collective excitations emerge from the long-wavelength fluctuation of the helicity angle $\gamma_p$ across the lattice, protected by the topology of the irreps space through the charge $Q_p\in\mathbb{Z}$. These modes, termed helicity phase modes or phasons, possess novel characteristics. The two concentric vortices per plaquette are coupled by interaction but not symmetry. Frustration affects only the outer vortex, resulting in the fragmentation of the excitation spectrum into a collective mode for the ordered fields and local excitations for the disordered components. The Mermin-Wagner theorem dictates the instability of ordered states to gapless magnons or phason modes, while disorder phases tend to order via thermal fluctuations according to the order-by-disorder paradigm \cite{Villain,Chandra,Chalker,Shenderbook}. Moreover, the VO/AO phases for $\pm D$ consist of different irreps, i.e., distinct conjugacy classes, that do not couple in the Hamiltonian. Hence, their phase boundary at $D=0$ signifies a topological phase transition associated with a spin liquid phase at the critical point, reminiscence of the deconfined critical point \cite{Senthil2004Science}. The CSL critical point can be extended by applying a magnetic field in the $z$-direction (see \textit{SM}\cite{SM}). Finally, transitions between ordered and fragmented phases, or within fragmented phases, offer intriguing avenues for studying non-Landau-type phase transitions.

\textit{{Acknowledgements:}} We thank Shivaji Sondhi for useful discussions. TD acknowledges research funding from the S.E.R.B. Department of Science and Technology, India, under Core Research Grant (CRG) Grant No. CRG/2022/00341 and acknowledges the computational facility at S.E.R.C. Param Pravega under NSM grant No. DST/NSM/R\&D HPC Applications/2021/39.

\vspace{1mm}
\bibliographystyle{apsrev4-2}
\bibliography{main.bib}

\begin{thebibliography}{59}
\bibitem [{dua()}]{dualspace}%
  \BibitemOpen
  \href@noop {} {}\bibinfo {note} {We define in the SM\cite{SM} the operation
  $\mathcal{T}$ for the dual vector as transpose $T$ followed by a
  normalization factor of $6/8$ to adjust the length of the plaquette spin to
  be $|\mathcal{S}_p|=6$.}\BibitemShut {Stop}%
\end{thebibliography}

\onecolumngrid
\newpage
\section{Supplementary Material}
\beginsupplement

\section{Detailed derivation of the Symmetry properties}
Here, we provide further details of the relevant mathematical constructions that are used in the main text. We start with a system of $\mathcal{N}$ spins. Much like how one starts in the quantum case with a direct product state basis to construct exotic entangled states, here we can also start with a many-body $3\mathcal{N}$-dimensional vector field as a direct sum basis: $\mathcal{S}=\oplus_i^{\mathcal{N}}{\bf S}_i$, where ${\bf S}_{i=1}\in \mathsf{O}(3)$. Then, the most general two-spin interaction Hamiltonian is written as $H=\mathcal{S}^T\mathcal{H}\mathcal{S}$, where $\mathcal{H}$ is the $3\mathcal{N}\times 3\mathcal{N}$ matrix-valued Hamiltonian. Short-range interaction and (discrete) translational symmetry drastically simplifies this Hamiltonian, giving a block-diagonal one. 

We assume that there exists a (conventional) unit cell with sublattices that are invariant under a point group symmetry $\mathsf{G}$. The spins sitting at the cell coordinates interact with the spins from the neighboring cells. This interaction term is translated back to a periodically equivalent interaction between the spins within the cell. This allows us to define a plaquette containing $n$ sublattices (counting the sites fully that are shared with the neighboring cells, and hence, the number of sublattices in a plaquette is larger than that in the primitive cell). In this prescription, the Hamiltonian $\mathcal{H}$ becomes block diagonal into a $3n\times 3n$ plaquette Hamiltonian $\mathcal{H}_p$, and the many-body spin vector field splits as $\mathcal{S}=\oplus_{p=1}^{N=\mathcal{N}/n}\mathcal{S}_p$, where $\mathcal{S}_p$ the vector field in the plaquette. 

Here, we focus on the square Kagome lattice, which has $n=8$ sites in a plaquette, giving a 24-dimensional reducible representation $\mathcal{S}_p$, as shown in Fig.1 (a), while the primitive unit cell has 6 spins  Therefore, the 'completeness' property of the plaquette spin turns out to be $\sum_{p=1}^N \mathcal{S}^T_p\mathcal{S}_p=8N$,  whereas $\mathcal{N}=6N$ is the total number of spin in the lattice of $N$ unit cell. To deal with this, we introduce a local weight factor $\eta_p$ in the definition of the dual vector, say, $\mathcal{S}_p^{\mathcal{T}}=\mathcal{S}_p^T\eta_p$, where $T$ corresponds to the transpose operator.  Then, the length of the vector is defined as $\mathcal{S}_p^{\mathcal{T}}\mathcal{S}_{p'}=\mathcal{S}_p^{T}\eta_p\mathcal{S}_{p'}=\eta_p\delta_{pp'}$. We approximate $\eta_p=6/8\mathcal{I}$ in each plaquette. The Monte Carlo result confirms that the obtained order parameters for phase phases are scaled with the group theory result by $6/8$.

Our first job is to find the irreducible representation of the Dihedral group $\mathsf{D}_4$ group in this vector field representation. The group elements are denoted by $\mathsf{D}_4=\{e, \mathsf{C}_4, \mathsf{C}_4^2, \mathsf{C}_4^3, \sigma_v^x, \sigma_v^{y}=\mathsf{C}_4^{-1}\sigma_v^{x}\mathsf{C}_4, \sigma_v^{xy}, \sigma_v^{yx}=\mathsf{C}_4^{-1}\sigma_v^{xy}\mathsf{C}_4\}$, where $\mathsf{C}_4$ is the four-fold rotation, $\sigma_v$ are the reflection with respect to the verticle plane passing through the $x,y-$ axis, or diagonal ($xy/yx$), as shown in Fig.~1(a).  In this $\mathcal{S}_p$-representation, we can split each of the $\mathsf{D}_4$ group elements as successive transformations on how the onsite spin ${\bf S}_i\in \mathsf{O}(3)$ undergoes an internal spin rotation, followed by how each component  $S^{\mu}_{i=1-8}$ of the  8 sublattices reorders in the plaquette vector $\mathcal{S}_p$. Noticeably further, the inner and outer squares of the square kagome lattice are decoupled from each other in terms of the $\mathsf{D}_4$ symmetries and give a trivial transformation between the two concentric squares of four sublattices.  In what follows, if we denote the $\mathcal{S}_p$-representation of the group elements $\mathsf{g}\in \mathsf{D}_4$ as $\mathcal{D}(\mathsf{g})$, then it can be decomposed into a direct product of three symmetries: $\mathcal{D}(\mathsf{g})=\mathcal{R}_I(\mathsf{g})\otimes \mathcal{R}_L(\mathsf{g})\otimes\mathcal{R}_S(\mathsf{g})$, where $\mathcal{R}_S(\mathsf{g})$ are the $3\times 3$ rotational matrices of the local $\mathsf{O}_i(3)$ spin, $\mathcal{R}_L(\mathsf{g})$ are the $4\times 4$ rotational matrices of the four sublattices, and $\mathcal{R}_I(\mathsf{g})$ is the $2\times 2$ transformation between the inner and outer squares. 

\begin{eqnarray}
\mathcal{D}(\mathsf{C}_4) &=& \left[\tau_0\otimes\mathcal{R}_L^{(1)}(\mathsf{C}_4)  + \tau_x\otimes\mathcal{R}_L^{(2)}(\mathsf{C}_4)\right] \otimes\mathcal{R}_S(\mathsf{C}_4),\nonumber\\ 
\mathcal{D}(\mathsf{C}^2_4)&=& \tau_x\otimes\mathbb{I}_{4 \times 4} \otimes \mathcal{R}_S(\mathsf{C}_4^2),\nonumber \\ 
\mathcal{D}(\mathsf{C}^3_4) &=& \left[\tau_0\otimes\mathcal{R}_L^{(2)}(\mathsf{C}_4)  + \tau_x\otimes\mathcal{R}_L^{(1)}(\mathsf{C}_4)\right]  \otimes \mathcal{R}_S(\mathsf{C}_4^3),\nonumber \\ 
\mathcal{D}(\sigma_v^{x}) &=& \left[\tau_0\otimes\mathcal{R}_L^{(1)}(\sigma_v^{x})  + \tau_x\otimes\mathcal{R}_L^{(2)}(\sigma_v^{x})\right]\otimes \mathcal{R}_S(\sigma_v^{x}),\nonumber \\
\mathcal{D}(\sigma_v^{y}) &=& \left[\tau_0\otimes\mathcal{R}_L^{(2)}(\sigma_v^{x})  + \tau_x\otimes\mathcal{R}_L^{(1)}(\sigma_v^{x})\right] \otimes \mathcal{R}_S(\sigma_v^{y}),\nonumber\\
\mathcal{D}(\sigma_v^{xy}) &=& \left[\tau_0\otimes\mathcal{R}_L^{(1)}(\sigma_v^{xy})  + \tau_x\otimes\mathcal{R}_L^{(2)}(\sigma_v^{xy})\right]  \otimes \mathcal{R}_S(\sigma_v^{xy}), \nonumber\\
\mathcal{D}(\sigma_v^{yx}) &=& \left[\tau_0\otimes\mathcal{R}_L^{(2)}(\sigma_v^{xy})  + \tau_x\otimes\mathcal{R}_L^{(1)}(\sigma_v^{xy})\right]  \otimes \mathcal{R}_S(\sigma_v^{yx}).
\end{eqnarray}
Here $\tau_0,\tau_x$ are Pauli matrices defining the internal symmetry $\mathcal{D}_4(\mathsf{g})$, and 
\begin{eqnarray}
\mathcal{R}_L^{(1)}(\mathsf{C}_4)  = \begin{pmatrix}
0 & 0 & 0 & 0\\
0 & 0 & 0 & 0\\
1 & 0 & 0 & 0\\
0 & 1 & 0 & 0\\
\end{pmatrix}, 
\mathcal{R}_L^{(2)}(\mathsf{C}_4)  = \begin{pmatrix}
0 & 0 & 1 & 0\\
0 & 0 & 0 & 1\\
0 & 0 & 0 & 0\\
0 & 0 & 0 & 0\\
\end{pmatrix}, 
\mathcal{R}_L^{(1)}(\sigma_v^x) = \begin{pmatrix}
0 & 0 & 0 & 0\\
0 & 0 & 0 & 0\\
0 & 0 & 0 & 0\\
0 & 0 & 0 & 1\\
\end{pmatrix},
\mathcal{R}_L^{(2)}(\sigma_v^x) = \begin{pmatrix}
0 & 0 & 1 & 0\\
0 & 1 & 0 & 0\\
1 & 0 & 0 & 0\\
0 & 0 & 0 & 0\\
\end{pmatrix}, \nonumber\\
\mathcal{R}_L^{(1)}(\sigma_v^{xy}) = \begin{pmatrix}
1 & 0 & 0 & 0\\
0 & 0 & 0 & 0\\
0 & 0 & 0 & 0\\
0 & 0 & 0 & 0\\
\end{pmatrix}, 
\mathcal{R}_L^{(2)}(\sigma_v^{xy}) = \begin{pmatrix}
0 & 0 & 0 & 0\\
0 & 0 & 0 & 1\\
0 & 0 & 1 & 0\\
0 & 1 & 0 & 0\\
\end{pmatrix}.
 \nonumber
\end{eqnarray}
Under $\mathsf{C}_4$, the continuous $\mathsf{O}_i(3)$ symmetry simply becomes a discrete angle of rotation by $2\pi/4$ with $L_z$ being the angular momentum, while under the mirror,  spin is rotated as an axial vector. This gives 
\begin{eqnarray}
\mathcal{R}_S(\mathsf{C}_4) = \begin{pmatrix}
0 & -1 & 0\\
1 & 0 & 0\\
0 & 0 & 1
\end{pmatrix}, ~~~
\mathcal{R}_S(\sigma _v^{x}) = \begin{pmatrix}
1 & 0 & 0\\
0 & -1 & 0\\
0 & 0 & -1
\end{pmatrix}, ~~~
\mathcal{R}_S(\sigma_v^{xy}) = \begin{pmatrix}
0 & -1 & 0\\
-1 & 0 & 0\\
0 & 0 & -1
\end{pmatrix},
\end{eqnarray}
and  $\mathcal{R}_S(\mathsf{C}_4^2) = (\mathcal{R}_S(\mathsf{C}_4))^2$,  $\mathcal{R}_S(\mathsf{C}_4^3) = (\mathcal{R}_S(\mathsf{C}_4))^3$,  $\mathcal{R}_S(\sigma _v^{y})=\mathcal{R}_S(\mathsf{C}_4)^{-1}\mathcal{R}_S(\sigma _v^{x})\mathcal{R}_S(\mathsf{C}_4)$, and $\mathcal{R}_S(\sigma_ v^{yx})=\mathcal{R}_S(\mathsf{C}_4)^{-1}\mathcal{R}_S(\sigma _v^{xy})\mathcal{R}_S(\mathsf{C}_4)$.

\subsection{Symmetry of the Hamiltonian}
The generic plaquette Hamiltonian is expressed in the main text as $H_p=\frac{1}{2}\mathcal{S}_p^\mathcal{T}\mathcal{H}_p\mathcal{S}_p$, where $\mathcal{H}_p$ is the $24\times 24$ {\it symmetric} matrix containing all possible nearest neighbor interactions. The symmetry constraints make many terms vanish or be identical to other terms.  Under a symmetry, the vector field transforms to  $\mathcal{S}_p^{\prime} = \mathcal{D}(\mathsf{g})\mathcal{S}$, $\forall \mathsf{g}\in \mathsf{D}_4$, and if the Hamiltonian to  $H_p$ is invariant, then the Hamiltonian matrix transforms as $\mathcal{D}^\mathcal{T}(\mathsf{g}) \mathcal{H}_p \mathcal{D}(\mathsf{g}) = \mathcal{H}_p$, $\forall p$. 

Under these conditions, we find that the interaction terms among the four triangles are related to each other by symmetry, while those within a triangle are independent of each other; see Fig.~1 (a).  Consider the one independent triangle at sites $i=\{0,1,2\}$ in Fig.~1 (a), and we obtain three distinct $3\times 3$ matrices between sites $i$ and $j$ : 
\begin{eqnarray}
    (\mathcal{H}_p)_{01} = \begin{pmatrix}
J^{xx} & D^{xy} & 0\\
D^{yx} & J^{yy} & 0\\
0 & 0 & J^{zz}
\end{pmatrix}, 
 (\mathcal{H}_p)_{12}=\begin{pmatrix}
J^{xx} & -D^{yx} & 0\\
-D^{xy} & J^{yy} & 0\\
0 & 0 & J^{zz}
\end{pmatrix}, \text{ and } 
 (\mathcal{H}_p)_{20}=\begin{pmatrix}
J'^{xx} & D'^{xy} & 0\\
-D'^{xy} & J'^{yy} & 0\\
0 & 0 & J'^{zz}
\end{pmatrix}. 
\end{eqnarray}
Therefore, we have nine independent parameters: three exchange interactions $J^{\mu\mu}$, $J'^{\mu\mu}$. and three DM interactions $D^{xy}$, $D^{yx}$, and  $D'^{xy}$. Due to in-plane inversion symmetry, no in-plane DM interaction is allowed. We take a simpler XXZ + DM interaction model in which $J^{\mu\mu}=J'^{\mu\mu}$, $J^{xx}=J^{yy}=J^{zz}/\Delta=J$, and $D^{xy}=\red{-}D^{yx}=D'^{xy}=JD$. This gives us three independent parameters, among which the global energy scaling by $J$ is removed, except its sign $\pm$ is considered in the main text. 

\subsection{Irreducible spin configurations}\label{sec:repGStates}
Finally, we find the irreducible representation of the $\mathcal{S}_p$ vector. There are five classes in the group $\mathsf{D}_4$ denoted by $E = \{\mathsf{e}\}, C_{4}= \{\mathsf{C}_4, \mathsf{C}_4^3\}, C_2 = \{\mathsf{C}_4^2\}, C_2' = \{\sigma_v^{xy}, \sigma_v^{yx}\}, C_2'' = \{\sigma_v^{x}, \sigma_v^y
\}$. The character table for this symmetry group is given in Table~\ref{tab:charac}.

\begin{table}[tb!]
    \centering    
\begin{tabular}{|c c c c c c c| } 
 \hline
 \hline
 $\mathsf{D}_{4}$ & $d_{\alpha}$ & $E$ & $2C_{4}$ & $2C_2''$ & $C_{2}$ & $2C_2'$ \\ [0.5ex] 
 \hline\hline
 $\mathsf{A}_{1}$ & 2 &  1 & 1 & 1 & 1 & 1\\ 
 \hline
 $\mathsf{A}_{2}$ & 4 & 1 & 1 & -1 & 1 & -1\\
 \hline
 $\mathsf{B}_1$ & 3 & 1 & -1 & 1  & 1 & -1\\
 \hline
 $\mathsf{B}_2$ & 3 & 1 & -1 & -1 & 1 & 1\\
 \hline
 $\mathsf{E}$ & 6 & 2 & 0 & 0 & -2 & 0\\ 
 \hline
 $\mathsf{S}_p$ & & 24 & 0 & -2 & 0 & -2 \\ 
 \hline
 \hline
\end{tabular}
\caption{Character table of the group $\mathsf{D}_4$. The last row corresponds to the characters of the reducible representation $\mathcal{S}_p$ for each class. N$_\textit{k}$C$_{\textit{k}}$ notion is used in the first row. N$_k$ is the number of elements in each conjugacy class, C$_k$.}
\label{tab:charac}
\end{table}

We have five irreps, which we denote by $m_{\alpha}$ for $\alpha=1-5$. Then the vector representation of the irreps is a direct sum of the irreps $\mathcal{M} = \bigoplus_{\alpha} d_{\alpha} m_{\alpha}$ with $d_{\alpha}$ giving the number of times the $\alpha$\textit{-th} irrep appears in the sum. $d_{\alpha}$ is calculated from orthogonality relation with the characters: $\chi_{m_{\alpha}}(C_\textit{k}), \chi_{\mathcal{M}}(C_\textit{k})$ of the 24-dimensional representations $m_{\alpha}(C_{k}), \mathcal{M}(C_{k})$ respectively, for each conjugacy class $C_k$, where $k$ runs over the five conjugacy classes: 
\begin{equation}
    d_{\alpha} = \frac{1}{h}\sum_{k} N_{k} \chi_{m_{\alpha}}(C_{k})
 \bar{\chi}_{\mathcal{M}}(C_\textit{k})
\end{equation}
where $h=8$ is the order of the group $\textbf{D}_4$, and $N_{\textit{k}}$ is the number of elements in $C_k$ conjugacy class. The values of $d_{\alpha}$ are given in the second column in Table \ref{tab:charac}.

\begin{figure}[tb!]
    \center
    \includegraphics[height=0.750\linewidth,width=0.750\linewidth]{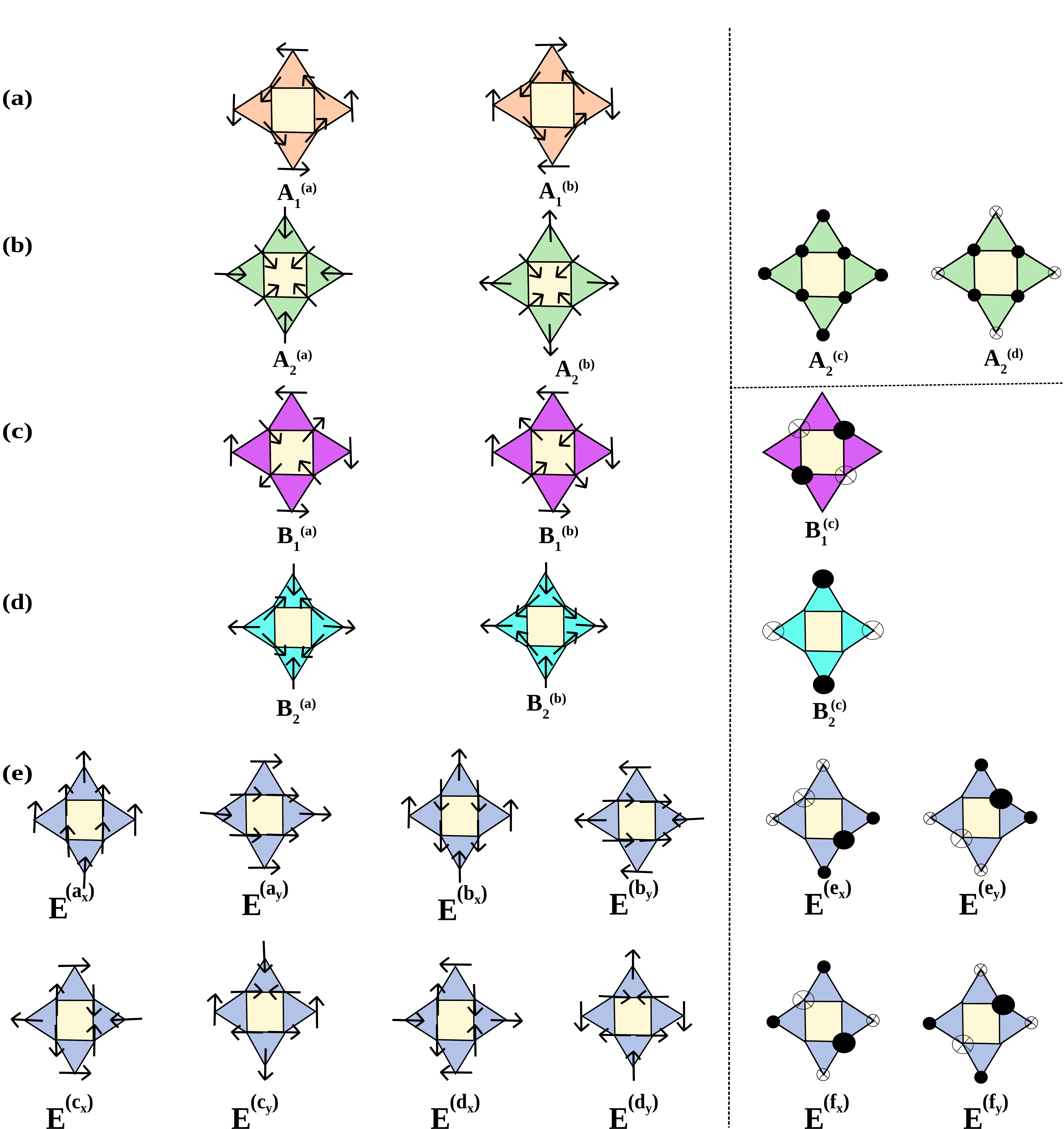}
    \caption{We plot all the irreps' basis functions. The verticle dashed line demarcates the out-of-plane irreps $\mathsf{m}_z$ on the right-hand side, among which only the top row satisfies the local constraint while the others do not. The horizontal arrows dive the spin direction for $S_i^{\perp}$, while the filled and open dots correspond to $S_i^z$. The size of the dots corresponds to $|S_i^z|$. For $\mathsf{A}_2^{\rm{(c,d)}}$,  the size of the dots is adjusted for $|S_i^z|=1$, while for $\mathsf{B}_{1,2}^{\rm{(c)}}$, sites with symbols give $|S_i^z|=\sqrt{2}$, while sites without symbols have $|S_i|=1$. Similar consideration is used for the $\mathsf{E}$ irreps that do not meet the local constraint. }
    \label{fig:repGStates}
  \end{figure}

The final task in this section is to find the basis functions $\mathcal{V}_{\alpha}$ of each irrep.  We denote the basis vectors as $|\mathcal{V}_{\alpha}^{\mu}\rangle$, where $\alpha=1$ for one-dimensional irreps, and $\mu=1,2$ (which are relabelled as $x,y$ in Fig.~\ref{fig:repGStates}) for the two-dimensional irrep $\mathsf{E}$. The basis vectors follow a relation : $\mathcal{D}(\mathsf{g})|\mathcal{V}_{\alpha}^{\mu}\rangle =\sum_{\mu'}(U_{\alpha}(\mathsf{g}))_{\mu\mu'}|\mathcal{V}_{\alpha}^{\mu'}\rangle$, $\forall \mathsf{g}$. $(U_{\alpha}(\mathsf{g}))_{\mu\mu'}$ are the $\mu\times\mu$-matrix for the $\mu$-dimensional irrep $\alpha$ defined for the group element $\mathsf{g}$. For the one-dimensional irreps $\mathsf{A}_{1,2}$ and $\mathsf{B}_{1,2}$, $U_{\alpha}(\mathsf{g})$ simply gives the character of the group, and then $|\mathcal{V}_{\alpha}^{\mu}\rangle$ are the simultaneous eigenvectors of the group elements with the character being the eigenvalue. They can be solved easily and the corresponding basis functions for the one-dimensional irreps are shown in Fig.~\ref{fig:repGStates}(a-d). For the two-dimensional $\mathsf{E}$ irrep, the orthogonal condition of the basis vector simplifies the above equation to $(U_{\alpha}(\mathsf{g}))_{\mu\mu'}=\langle\mathcal{V}_{\alpha}^{\mu}|\mathcal{D}(\mathsf{g})|\mathcal{V}_{\alpha}^{\mu'}\rangle$. We solve this matrix for the $\mathsf{E}$ irrep for each group elements, which comes out to be  $U_{\mathsf{E}}(\mathsf{e})=\mathbb{I}_{2\times 2}$, $U_{\mathsf{E}}(\mathsf{C}_4)=-i\tau_y$, $U_{\mathsf{E}}(\mathsf{C}_4^2)=-\mathbb{I}_{2\times 2}$, $U_{\mathsf{E}}(\mathsf{C}_4^3)=i\tau_y$, $U_{\mathsf{E}}(\mathsf{\sigma}_v^x)=\tau_z$, $U_{\mathsf{E}}(\mathsf{\sigma}_v^y)=-tau_z$, $U_{\mathsf{E}}(\mathsf{\sigma}_v^{xy})=\tau_x$, $U_{\mathsf{E}}(\mathsf{\sigma}_v^{yx})=-\tau_x$.  $\tau_{\mu}$ are the $2\times 2$ Pauli matrices.

We have the multiplets as $d_{\alpha}=2,4,3,3$ for the four one-dimensional irreps $\mathsf{A}_1$, $\mathsf{A}_2$, $\mathsf{B}_1$, $\mathsf{B}_2$, giving 12 basis vectors, while the two-dimensional irrep with multiplicity $d_{\mathsf{E}}=6$ gives another 12 basis vectors, as shown in Fig.~\ref{fig:repGStates}(e).  Among them, sixteen are in-plane, defined in the set $\mathsf{m}_{\perp}$, and eight are out-of-plane, defined in the set $\mathsf{m}_z$ in the main text. Among them, six out-of-plane irreps do not satisfy the local constraint of $S=1$ per site. 

\subsection{XXZ and DM interactions}

In the plaquette Hamiltonian, after substituting $\mathcal{S}_p=\sum_{\alpha=1}^{3n} m_{\alpha}\mathcal{V}_{\alpha}$, we obtain a Hamiltonian that is block diagonal between the irreps but contains cross-terms along the multiplicity within an irrep. So we define a $d_{\alpha}$-dimensional spinor field for each irrep as $\boldsymbol{{m}}_{\alpha}:=({m}_{\alpha}^{(1)}~...~ {m}_{\alpha}^{(d_{\alpha})})^{T}\in \mathsf{O}_p(d_{\alpha})$, in which the plaquette Hamiltonian splits as
\begin{eqnarray}
    H_p=\sum_{\alpha=1}^{5} \boldsymbol{{m}}_{\alpha}^T \mathcal{H}_{\alpha}\boldsymbol{{m}}_{\alpha},
\end{eqnarray}
where we have suppressed the plaquette index on the right-hand side. $\mathcal{H}_{\alpha}$ is a $d_{\alpha}\times d_{\alpha}$ matrix. The $\mathsf{O}_p(d_{\alpha})$ symmetry of each irrep breaks into $\mathsf{O}_p(2)$ and $\mathsf{Z}_2$ symmetry as follows. 

For $\alpha=1$, the $\mathsf{A}_1$ irrep with $d_{1}=2$  multiplets follows an $\mathsf{O}_p(2)$ symmetry. 

For $\alpha=2$,  the $\mathsf{A}_2$ irrep with $d_2=4$, we have an emergent $\mathsf{O}_p(2)\times \mathsf{O}_p(2)$ symmetry among the multiplets, giving $\mathcal{H}_{\mathsf{A}_2}=\mathcal{H}_{\mathsf{A}_2^{\rm{(a,b)}}}\oplus \mathcal{H}_{\mathsf{A}_2^{\rm{(c,d)}}}$. This is obvious because $\mathsf{A}_{2}^{\rm{(a,b)}}$ consists of coplanar spins while $\mathsf{A}_{2}^{\rm{(c,d)}}$ are the two out-of-plane spins. 

For both $\alpha=3,d$, the $\mathsf{B}_{1,2}$ irreps with $d_{3,4}=3$, we have an emergent $\mathsf{O}_p(2)\times \mathsf{Z}_2$ symmetry with $\mathcal{H}_{\mathsf{B}_{1,2}}=\mathcal{H}_{\mathsf{B}_{1,2}^{\rm{(a,b)}}}\oplus \mathcal{H}_{\mathsf{B}_{1,2}^{\rm{(c)}}}$. Here, the $\mathsf{B}^{\rm{(a,b)}}_{1,2}$ multiplets are coplanar spins forming  $\mathsf{O}(2)$ symmetry, while  $\mathsf{B}^{\rm{(c)}}_{1,2}$ consists of out-of-plane spins that do not obey local constraints. 

For $\alpha=5$, the two-dimensional $\mathsf{E}$ irrep with $d_5=6$, each component of each multiplicity gives emergent  $\mathsf{O}_p(2)$ rotation as $\mathcal{H}_{\mathsf{E}}=\mathcal{H}_{\mathsf{E}^{\rm{(a,b)}}}\oplus \mathcal{H}_{\mathsf{E}^{\rm{(c,d)}}}\oplus \mathcal{H}_{\mathsf{E}^{\rm{(e,f)}}}$. 

All the $\mathsf{O}_p(2)$ invariant $2\times 2$ Hamiltonian matrices for all irreps have this general form
\begin{eqnarray}
(\mathcal{H}_{\alpha})_{k,k'} = \epsilon_{\alpha}^{(k+)}\sigma_0 + \epsilon_{\alpha}^{(k-)}\sigma_z + \lambda_{\alpha}^{(kk')}\sigma_x,
\end{eqnarray}
where $k,k'=1,2\in$ (a,b) or (c,d) or (e,f), and $\epsilon^{k\pm}_{\alpha}=[\epsilon^{(k)}_{\alpha}\pm \epsilon^{(k')}_{\alpha}]/2$ and $\epsilon^{(k)}_{\alpha}$ is the onsite energy for the $k^{\rm th}$ multiplet of the $\alpha$-irrep, and $\lambda_{\alpha}^{(kk')}$ is the `hopping energy' between the $k$ and $k'$ multiples. The onsite energies of the two vortices with different helicities are  $\epsilon_{\mathsf{A}_1^{\rm(a)}}=\epsilon_{\mathsf{A}_2^{\rm(a)}}=2\sqrt{2}+2(\sqrt{2}-1)D$,  $\epsilon_{\mathsf{A}_1^{\rm(b)}}=\epsilon_{\mathsf{A}_2^{\rm(b)}}=-2\sqrt{2}-2(\sqrt{2}+1)D$, while the energy cost to change the helicity angle is $\lambda_{\mathsf{A}_1^{\rm(a,b)}}=\lambda_{\mathsf{A}_2^{\rm(a,b)}}=-4D$. The same for the two anti-vortices are:
$\epsilon_{\mathsf{B}_1^{\rm(a)}}=\epsilon_{\mathsf{B}_2^{\rm(a)}}=-2\sqrt{2}+2(\sqrt{2}+1)D$,  $\epsilon_{\mathsf{B}_1^{\rm(b)}}=\epsilon_{\mathsf{B}_2^{\rm(b)}}=2\sqrt{2}-2(\sqrt{2}-1)D$, $\lambda_{\mathsf{B}_1^{\rm(a,b)}}=\lambda_{\mathsf{B}_2^{\rm(a,b)}}=-4D$. The out-of-plane irreps with parallel and anti-parallel spins and spin-flip energies between them as $\epsilon_{\mathsf{A}_2^{\rm(c)}}=6\Delta$, $\epsilon_{\mathsf{A}_2^{\rm(d)}}=-2\Delta$,  $\lambda_{\mathsf{A}_2^{\rm(c,d)}}=4\Delta$. The two irreps with only inner and out-square out-of-plane spins have the onsite energy: $\epsilon_{\mathsf{B}_1^{\rm(c)}}=\epsilon_{\mathsf{B}_2^{\rm(c)}}=-4\Delta$. Each two-dimensional irreps is degenerate. The in-plane FM $\mathsf{E}$ irreps have the energies $\epsilon_{\mathsf{E}^{\rm(a)}}=6$, $\epsilon_{\mathsf{E}^{\rm(b)}}=-2$, and  their hopping energy $\epsilon_{\mathsf{E}^{\rm(a,b)}}=4$. The in-plane AFM $\mathsf{E}$ irreps have the energies $\epsilon_{\mathsf{E}^{\rm(c)}}=4D-2$, $\epsilon_{\mathsf{E}^{\rm(d)}}=-4D-2$, and  $\epsilon_{\mathsf{E}^{\rm(c,d)}}=-4$. The two out-of-plane $\mathsf{E}$ irreps that do not mix have the energies $\epsilon_{\mathsf{E}^{\rm(e)}}=2\sqrt{2}$, $\epsilon_{\mathsf{E}^{\rm(b)}}=-2\sqrt{2}$. All energies are multiplied with $J$. 

The explicit form of Hamiltonian in terms of the matrix elements in  the basis of the irrep order parameter is 
\begin{equation}{\label{eq:xxzOp}}
\begin{split}
    H_p = \sum_{\alpha=\mathsf{A}_{1,2},\mathsf{B}_{1,2}}\sum_{k,k'} (\mathcal{H}_{\alpha})_{k,k'}m_{\alpha}^{(k)}m_{\alpha}^{(k')} +  \sum_{k,k'} (\mathcal{H}_{\mathsf{E}})_{k,k'}{\bf m}_{\mathsf{E}}^{(k)}\cdot {\bf m}_{\mathsf{E}}^{(k')} + \sum_{\alpha=\mathsf{B}_{1,2},k={\rm c}}(\mathcal{H}_{\mathsf{E}})_{k,k}(m_{\alpha}^k)^2.
\end{split}
\end{equation}
where $k,k'=$a,b for all irreps, and in addition, we have  $k,k'=$c,d for $\mathsf{A}_2$ and $k,k'=$c,d, and $k,k'=$e,f for $\mathsf{E}$.

Then, for all $\mathsf{O}_p(2)$ order parameters, diagonalize the corresponding $2\times 2$  $\mathcal{H}_{\alpha}$ matrices by the orthogonal transformation: 

$$ \begin{pmatrix}
    \tilde{m}_{\alpha}^{(k)}\\
    \tilde{m}_{\alpha}^{(k')}
\end{pmatrix} 
 = \left[\sigma_0\cos \phi_{\alpha}^{(k,k')}-i\sigma_y \sin\phi_{\alpha}^{(k,k')}\right]
\begin{pmatrix}
   m_{\alpha}^{(k)}\\
   m_{\alpha}^{(k')}
\end{pmatrix}$$
where $\phi_{\alpha}^{(k,k')}$ is a fixed angle of rotation that diagonalizes the corresponding irrep multiplets. Eventually, we obtain a fully diagonal Hamiltonian as
\begin{equation}{\label{eq:xxzdiag}}
    H_p = \sum_{\nu=(\alpha,k=1,d_{\alpha})} E_{\nu}|\tilde{m}_{\nu}|^2.
\end{equation}
We have abandoned the $\alpha$ and $k$ symbols for the irreps and multiplicity and combined them into a single symbol $\nu$ which runs from 1 to $3n$ in the eigenmodes, for simplicity. 
Here $E_{\nu} = \epsilon_{\alpha}^+\pm \sqrt{(\epsilon_{\alpha}^-)^2 + \lambda_{\alpha}^2}$ for each $\mathsf{O}_p(2)$ multipltes of $\alpha$-irreps.  Their explicit forms are
\begin{eqnarray}
E_{\nu=1,2} &=& -2D\pm 2\sqrt{D^2+(1+D)^2},\qquad \rm{for}~\alpha=\mathsf{A}_{1}^{\rm(a,b)},\nonumber\\
E_{\nu=3,4}&=&E_{\nu=1,2}, \qquad \text{\hspace{2.9cm}}  \rm{for}~\alpha=\mathsf{A}_{2}^{\rm(a,b)},\nonumber\\
E_{\nu=5,6} &=& 2\Delta (1 \pm \sqrt{5}), \qquad \text{\hspace{2.1cm}} \rm{for}~\alpha=\mathsf{A}_{2}^{\rm(c,d)},\nonumber\\
E_{\nu=7,8} &=& 2D \pm 2\sqrt{D^2 + 2(1-D)^2},
\text{\hspace{0.8cm}} \rm{for}~\alpha=\mathsf{B}_{1}^{\rm(a,b)},\nonumber\\
E_{\nu=9} &=& -4\Delta, 
\text{\hspace{3.9cm}} \rm{for}~\alpha=\mathsf{B}_{1}^{\rm(c)},\nonumber\\
E_{\nu=10-12}&=&E_{\nu=7-9}
\text{\hspace{3.6cm}} \rm{for}~\alpha=\mathsf{B}_{2}^{\rm(a,b,c)},\nonumber\\
E_{\nu=13,14} &=&2\pm 2\sqrt{5},  
\text{\hspace{3.3cm}} \rm{for}~\alpha=\mathsf{E}^{\rm(a,b)},\nonumber\\
E_{\nu=15,16}&=& -2\pm 2\sqrt{1+4D^2}, \text{\hspace{1.9cm}} \rm{for}~\alpha=\mathsf{E}^{\rm(c,d)},\nonumber\\
E_{\nu=17,18}&=&\pm 2\sqrt{2}\Delta \text{\hspace{3.5cm}} \rm{for}~\alpha=\mathsf{E}^{\rm(e,f)}.
%
\end{eqnarray} 
All the energies are defined with respect to $J$. The values of the angle $\phi$ are:  \\
\begin{eqnarray}
    \phi_{\mathsf{A}_1^{\rm (a,b)}} &=& \frac{1}{2}\tan^{-1}\left(\frac{D}{\sqrt{2}(1+D)}\right) , \text{\hspace{0.51cm}} \phi_{\mathsf{A}_2^{\rm (a,b)}} = \phi_{\mathsf{A}_1^{\rm (a,b)}}, \text{\hspace{0.51cm}} \phi_{\mathsf{A}_1^{\rm (c,d)}} = -\frac{1}{2}\tan^{-1}\left(\frac{1}{2}\right),\nonumber \\
    \phi_{\mathsf{B}_1^{\rm (a,b)}} &=& \frac{1}{2}\tan^{-1}\left(\frac{D}{\sqrt{2}(1-D)}\right), \text{\hspace{0.51cm}} \phi_{\mathsf{B}_2^{\rm (a,b)}} = \frac{1}{2}\tan^{-1}\left(\frac{-D}{\sqrt{2}(1-D)}\right),\nonumber\\
    \phi_{\mathsf{E}^{\rm (a,b)}} &=& -\frac{1}{2}\tan^{-1}\left(\frac{1}{2}\right), \text{\hspace{0.51cm}} \phi_{\mathsf{E}^{\rm (c,d)}} = \frac{1}{2}\tan^{-1}\left(\frac{1}{2D}\right).
\end{eqnarray}

\section{Details of Classical Monte Carlo}
In the classical Monte Carlo calculation, the final temperature is achieved by annealing from the high temperature at each step with $8 \times 10^{5}$ Monte Carlo steps. The expectation values of the observables are calculated by taking the average over the last $7 \times 10^{5}$ configurations of a total $8 \times 10^{5}$ Monte Carlo steps with system size $N=6L^2$ considering periodic boundary conditions, with $L$ number of unit cells. All the static structure factor averages are performed over system size, $L=20$ at temperature $10^{-3}$. The position vectors of each sublattice (denoted with indices 0,1, ... in Fig.~1(a) of main text) are taken as considering the origin at the center of the square, 
 \begin{eqnarray}
\delta_0 =  \left(\frac{-1}{4},\frac{-1}{4} \right), \text{\hspace{0.251cm}} \delta_1 =  \left(\frac{1}{4},\frac{-1}{4} \right), \text{\hspace{0.251cm}}
\delta_2 =  \left(\frac{1}{4},\frac{1}{4} \right), \text{\hspace{0.251cm}} 
\delta_3 =  \left(\frac{-1}{4},\frac{1}{4} \right), \text{\hspace{0.251cm}}
\delta_4 =  \left(0,\frac{-1}{2} \right), \text{\hspace{0.251cm}} \text{\hspace{0.251cm}} \delta_5 =  \left(\frac{1}{2},0 \right)
 \end{eqnarray}

\begin{figure}[htb!]
    \centering
    \includegraphics[height=0.25\linewidth,width=0.65\linewidth]{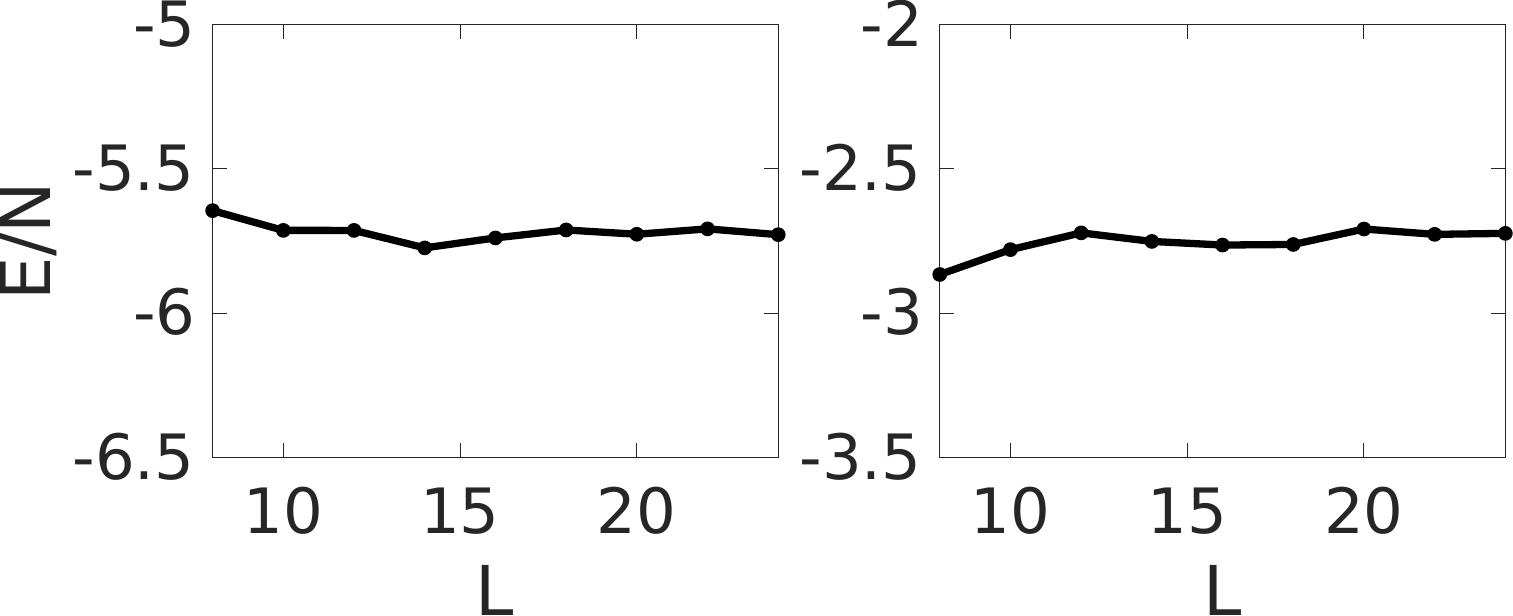}
    \caption{We plot the total energy per site vs the system size along one axis, L. The number of sites in the lattice is N=6*L*L. Left: The ordered state with D=-3.0 and $\Delta=0.0$; Right: The spin liquid phase at D=0.0 and $\Delta=4.00$.}
    \label{fig:FinitesizeFig}
\end{figure}
 
\section{Structure Factor Plots}
In this section, we list the real space spin configurations of all the phases and their respective structure factors. As defined in the main text, the different structure factors are  
\begin{eqnarray}
    \chi(\textbf{k}) &=& 1/\mathcal{N} \sum_{i,j} \langle \textbf{S}_i \cdot \textbf{S}_j \rangle \exp{({{\rm i}\textbf{k} \cdot (\textbf{r}_i - \textbf{r}_j)})} \nonumber \\
    \chi^{\perp}(\textbf{k}) &= & 1/\mathcal{N} \sum_{i,j} \langle \textbf{S}^{\perp}_i \textbf{S}^{\perp}_j \rangle \exp{({{\rm i}\textbf{k} \cdot (\textbf{r}_i - \textbf{r}_j)})} \nonumber \\
    \chi^{z}(\textbf{k}) &=& 1/\mathcal{N} \sum_{i,j} \langle \textbf{S}^{z}_i \textbf{S}^{z}_{j} \rangle \exp{({{\rm i}\textbf{k} \cdot (\textbf{r}_i - \textbf{r}_j)})} 
\end{eqnarray}

\begin{figure}[tb!]
    \center
    \includegraphics[height=1.0\linewidth,width=0.90\linewidth]{Phases_SupplyMat_OrderParam.pdf}
    \caption{The real spin configurations (left panel) and the corresponding structure factor (middle panel) are plotted for various phases for the AFM coupling $J=+1$. The ensemble of order parameters, which are mentioned in the main text, for each phase is presented in the right panel. (a) Order phase (red region in the phase diagram) with staggered anti-vortices between the neighboring sites, showing Bragg-like peaks at a finite but preferential wavevector. (b) Mixed or fragmented phase where the inner anti-vortices turn into an AFM-anti-vortex with opposite $S_i^z$ components, while $S_i^z=0$ for the outer anti-vortex. The $S_i^z$ values, however, take random values and show disorder features in the corresponding structure factor without any pinch-point correlation. This is expected as the inner vortices become decoupled from each other, lacking any significant correlation between them. (c) A CSL phase (close to the $\mathsf{Z}_2$ CSL phase) showing larger spectral weight the $S_i^z$ correlation function with pinch-points. (d) The mixed or fragmented phase for $D<0$ is similar to the mixed phase for $D>0$, except here, vortices replace the anti-vortices. (e) Orderd phase for D<0, similar to the $D>0$ case in (a), with vortices replacing anti-vortices. (f) A collinear out-of-plane FM phase arising in the limit of strong our-of-plane anisotropy term $\Delta\rightarrow -\infty$.  }
    \label{fig:Phases_Antiferro}
  \end{figure}

\begin{figure}[tb!]
    \center
    \includegraphics[height=0.5750\linewidth,width=0.920\linewidth]{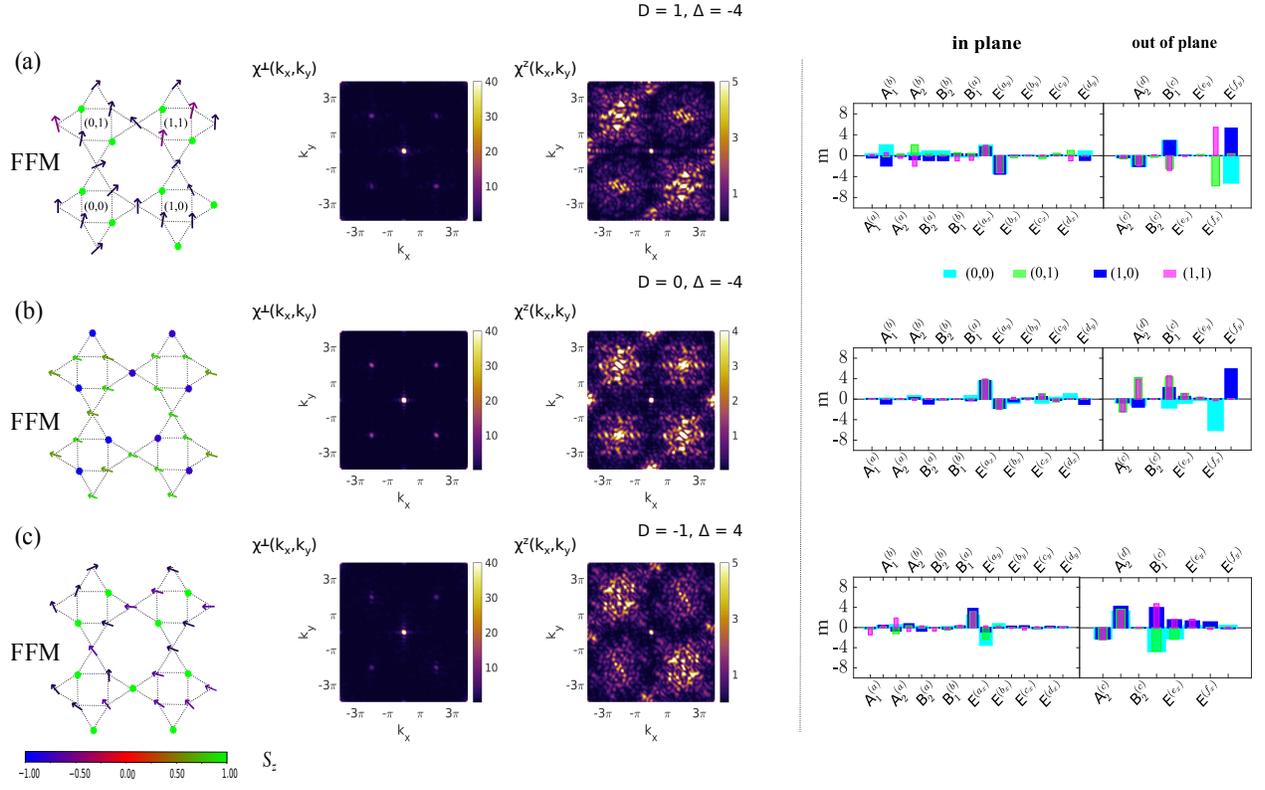}
    \caption{Similar to Fig. S2, but for the FM interaction $J=-1$. All three phases shown here are the fragmented phases at different values of $D$ and $\Delta$, showing pinch-point in the $S_i^z$ correlation function, but FM ordering in the in-plane component. }
    \label{fig:Phases_ferro}
  \end{figure}

\section{Soft-spin Approximation}\label{sec:soft}
\begin{figure}[htb!]
    \centering
    \includegraphics[height=0.56\linewidth,width=0.75\linewidth]{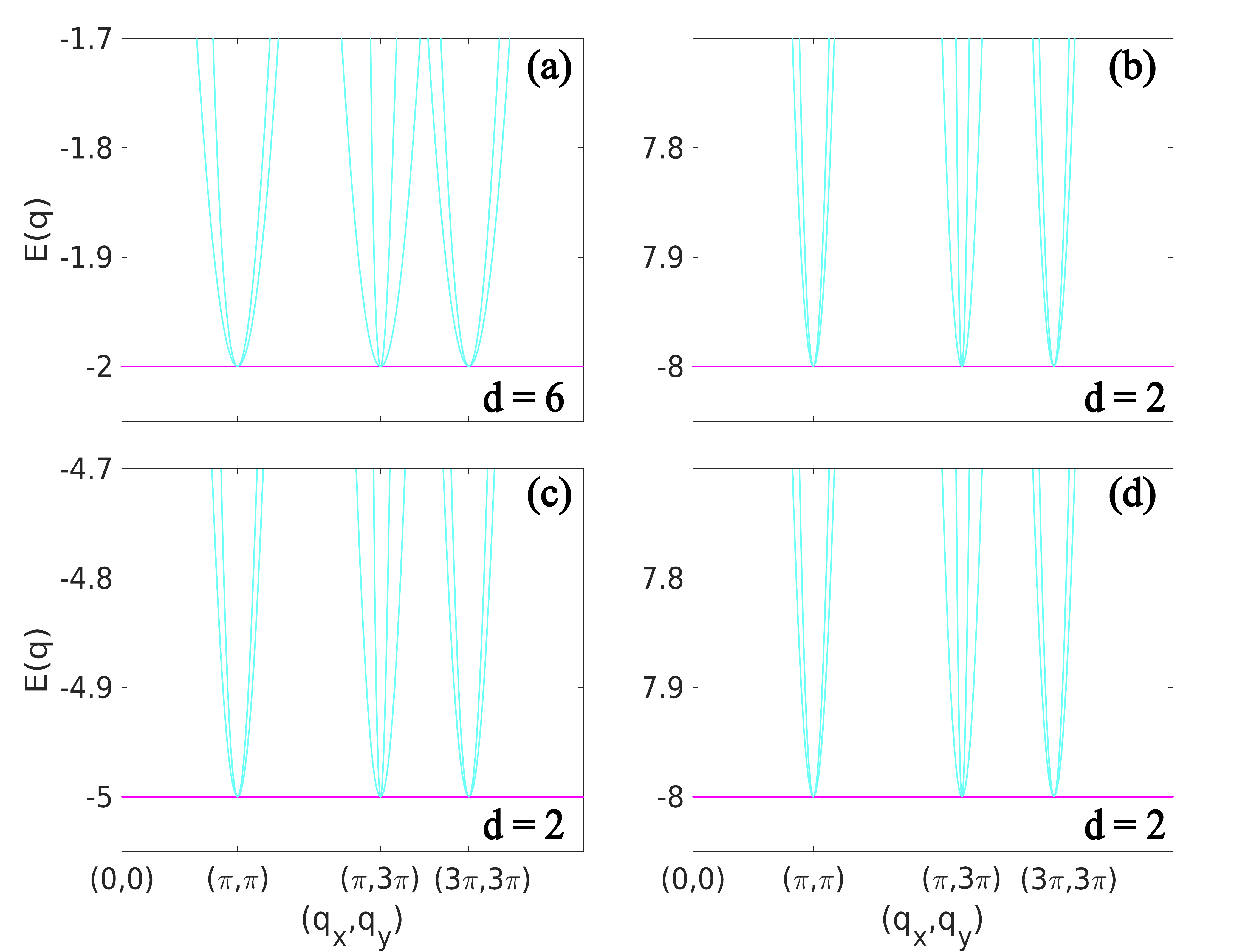}
    \caption{Energy dispersion of the Hamiltonian $\mathcal{H}({\bf q})$  at four 
    with re)spective degeneracy of flat bands d, for (a) $\Delta = 1.0, D = 0.0$ (CSL), (b) $\Delta = 4.0, D= 1.0$ (Mixed phase) for $J=+1$ and (c) $\Delta = -2.5, D = 0.0$ (d) $\Delta = 4.0, D = 1.0$ (Mixed phases) for J = -1}
    \label{fig:Dispersions}
  \end{figure}

In this section, we analyze the Hamiltonian in Eq.~\ref{eq:Hamxxz} with 'soft spin' approximation i.e. spin length constraint (|\textbf{S$_i$}|$^2=1$) is softened from exact value of 1 per site to the global value of $\sum_{i}^N|\textbf{S}_i|^2 = NS$. Because of the global constraint, we have a uniform (fixed) chemical potential (Lagrangian multiplier) in the theory. Then, following Ref.~\onlinecite{Yan2023Clas_Detailed}, we have diagonalized the Hamiltonian in the Fourier space of the spin. There, a spin vector is defined per unit cell, not in the plaquette, which means we have six sublattices as $\mathcal{S}_i$=(S$_0^x$, S$_0^y$, S$_0^z$, S$_1^x$, ..., S$_5^z$). We Fourier transform the spin vector as $\mathcal{S}(\textbf{q}) = \frac{1}{\sqrt{N}}\sum_i \mathcal{S}_i e^{-i \textbf{q}\cdot \textbf{r}_i}$, where $\textbf{r} = a \textbf{n}_1 + b \textbf{n}_2$ with integers a, b and unit vectors $\textbf{n}_1 = (1,0), \textbf{n}_2 = (0,1)$. The Hamiltonian is then diagonal in the momentum space as 
\begin{equation}
    H =  \sum_{\bf q}\mathcal{S}^T(\textbf{q})^{T} \mathcal{H}(\textbf{q}) \mathcal{S}(\textbf{q}),
\end{equation}
where $\mathcal{H}(\textbf{q})$ is a $18\times 18$ matrix. We can now diagonalize the $\mathcal{H}(\textbf{q})$ matrix, which gives the energy eigenvalues $E_{\nu}({\bf q})$. The lowest energy state is the ground state, and then we plot a few low-energy excited states in Fig.~\ref{fig:Dispersions}. 

We note that the analysis on the Fourier basis leads to a violation of the local constraint and hence, inconsistency is expected between the real-space model and the Fourier space one, especially in the spin liquid phase. In the CSL phase, we find an extremely flat band as the lowest energy state, suggesting extensive degeneracy as expected here. We see the flat band in all the mixed phases as well. In addition, the spectrum is gapless in both phases, with gapless points present at $(\pm\pi,\pm\pi), (\pm\pi,\pm 3\pi)$, and $(\pm3\pi,)\pm3\pi)$, as shown in Fig.~\ref{fig:Dispersions}. The band degeneracy, denoted with $d$ in the spectrum at each region is different: $d$=4(2) for $\Delta<1 (>1)$, $d$=6 at $\Delta=1$ in the CSL phase where $D$=0; and $d$=2 for mixed phases both for $J$ = +1 and -1. Hence, there is no simple positive sum of the constrainer rule here; the direct matching of singular/non-singular bands to emergent gauge fields/fragility is not possible. 

As discussed rigorously in the main text, the spin liquids (cyon(/black) colored phase for J=+1(/-1)) phase has pinch points belonging to the algebraic class of CSLs with 'emergent' low-energy gauge field excitations. The mixed (black-colored phase for $J$=+1) phase has no pinch points and, hence, belongs to the fragile class of CSLs. All the other ordered phase regions have dispersive bands.

\subsection{Finite Magnetic field}\label{sec:field}

\begin{figure}[htb!]
    \centering
    \includegraphics[height=0.25\linewidth,width=0.5\linewidth]{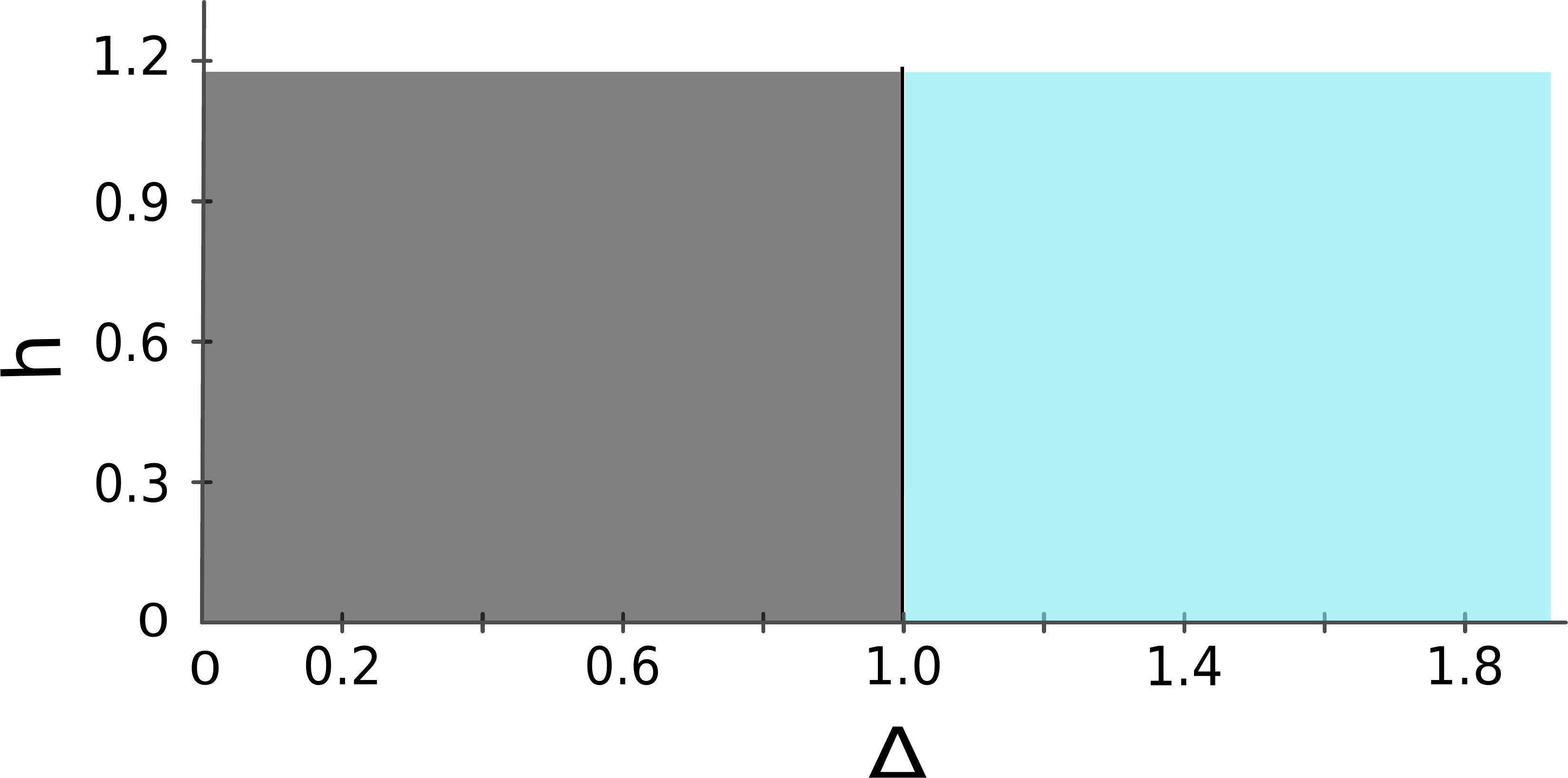}
    \caption{Phase diagram at D=0, as a function of h and $\Delta$. For $\Delta<1$, the phase is a mixed phase and spin liquid for another case. The mixed phase here is unstable for any finite value of D; the phase becomes ordered in and out-of-plane for non-zero D value.}
    \label{fig:FieldPhases}
  \end{figure}
  
 The external magnetic field is applied along the $z$-axis to the Hamiltonian, now written as 
\begin{equation}
    H_{mag} = H_{\rm XXZ-DM} - h \sum_{i} S_i^z.
\end{equation}
The phase diagram as a function of $h$ and $\Delta$ is presented in Fig.~\ref{fig:FieldPhases} for $D=0$. A mixed phase of disordered in-plane spins with ordered out-of-plane components is observed at $D=0$ for $\Delta<1$ with increasing $h$. The in-plane disordered spins exhibit a coexisting Bragg-like leak at $(0,4\pi)$, and pinch points at $(\pm \pi,\pm 3\pi)$. The ordering along the $z$ components is FM type. This phase is unstable for any finite value of $D$. A finite value of $D$ gives an ordered phase depending on the sign of the $D$ value, where the in-plane spins form an ordered supercell structure and the out-of-plane spins are ferromagnetically ordered. As $\Delta>1$, the spins become disordered both in in-plane and out-of-plane components. This phase also has pinch-points in the correlation function, indicating power-law correlations. This phase survives at finite values of $D$. Therefore, we conclude that, by applying the external magnetic field, the spin liquid phase can be stabilized in these materials.   

\end{document}